\documentclass[preprint,12pt]{elsarticle}

\usepackage{amssymb}
\usepackage{amsmath}

\usepackage{lineno}

\usepackage{booktabs}
\usepackage[final]{changes}
\usepackage{cleveref}
\usepackage{gensymb}
\usepackage[acronym,toc,nogroupskip]{glossaries}
\usepackage{multirow}
\usepackage{pgfplots}
\pgfplotsset{compat=1.17}
\usepgfplotslibrary{groupplots}
\usepackage{siunitx}
\usepackage{tabularx}
\usepackage{tikz}
\usetikzlibrary{patterns,arrows,shapes}
\usepackage{subcaption}
\usepackage{todonotes}
\usepackage{url}
\usepackage[T1]{fontenc}

\newacronym{bte}{BTE}{boundary traction element}
\newacronym{cass}{CASS}{continuous Airy-based for stress-singularities}
\newacronym{da}{DA}{digital annealing}
\newacronym{dcem}{DCEM}{deep complementary energy method}
\newacronym[plural=DOFs, longplural=degrees of freedom]{dof}{DOF}{degree of freedom}
\newacronym{fem}{FEM}{finite element method}
\newacronym{hre}{HRE}{higher-order rectangular elements}
\newacronym{hsf}{HS-F}{hybrid stress-function}
\newacronym{iga}{IGA}{isogeometric analysis}
\newacronym{nurbs}{NURBS}{non-uniform rational B-Splines}
\newacronym{pde}{PDE}{partial differential equation}
\newacronym{pinn}{PINN}{physics-informed neural network}
\newacronym{qa}{QA}{quantum annealing}
\newacronym{sa}{SA}{simulated annealing}

\begin{document}
\newcommand{\lp}{\left(}
\newcommand{\rp}{\right)}
\newcommand{\lb}{\left[}
\newcommand{\rb}{\right]}
\newcommand{\lcb}{\left\{}
\newcommand{\rcb}{\right\}}

\newcommand{\x}[1][]{x_{#1}}

\newcommand{\strainSymbol}{\varepsilon}
\newcommand{\strainTensor}{\boldsymbol{\strainSymbol}}
\newcommand{\strainTensorComponent}[1]{\strainSymbol_{#1}}

\newcommand{\stressSymbol}{\sigma}
\newcommand{\stressTensor}{\boldsymbol{\stressSymbol}}
\newcommand{\stressTensorRef}{{\boldsymbol{\hat{\stressSymbol}}}}
\newcommand{\coeffsAnsatz}{\boldsymbol{c}}
\newcommand{\stressTensorComponent}[1]{\stressSymbol_{#1}}
\newcommand{\stressTensorComponentRef}[1]{\stressSymbol^{\text{ref}}_{#1}}
\newcommand{\stress}{\stressSymbol}

\newcommand{\density}{\rho}
\newcommand{\youngsModulus}{E}
\newcommand{\poissonsRatio}{\nu}

\newcommand{\displacementSymbol}{u}
\newcommand{\displacement}{\displacementSymbol}
\newcommand{\displacementPrescribed}{\hat{\displacement}}
\newcommand{\displacementVector}{\boldsymbol{\displacementSymbol}}
\newcommand{\displacementVectorPrescribed}{\boldsymbol{\displacementPrescribed}}
\newcommand{\tractionSymbol}{t}
\newcommand{\traction}{\tractionSymbol}
\newcommand{\tractionVector}{\boldsymbol{\tractionSymbol}}
\newcommand{\tractionPrescribed}{\hat{\traction}}
\newcommand{\tractionVectorPrescribed}{\hat{\tractionVector}}
\newcommand{\bodyForceDensity}{f}

\newcommand{\domain}{\Omega}
\newcommand{\boundary}{\Gamma}
\newcommand{\boundaryDisp}{\boundary^{\displacement}}
\newcommand{\boundaryTraction}{\boundary^{\stressSymbol}}
\newcommand{\normal}{n}
\newcommand{\normalVector}{\boldsymbol{n}}

\newcommand{\intDomain}[1]{\int_{\domain} #1\,\text{d}\domain}
\newcommand{\intBoundaryDisp}[1]{\int_{\boundaryDisp} #1\,\text{d}\boundary}

\newcommand{\complianceTensorSymbol}{S}
\newcommand{\complianceTensor}{\boldsymbol{\complianceTensorSymbol}}
\newcommand{\complianceTensorComponent}[1]{\complianceTensorSymbol_{#1}}
\newcommand{\complianceMatrix}{\mathcal{\complianceTensorSymbol}}

\newcommand{\superscriptComplementaryEnergy}{*}
\newcommand{\totalComplEnergy}{\Pi^{\superscriptComplementaryEnergy}}

\newcommand{\complStrainEnergy}{U^{\superscriptComplementaryEnergy}}
\newcommand{\complStrainEnergyDensity}{\complStrainEnergy_0}
\newcommand{\complExternal}{W^{\superscriptComplementaryEnergy}}
\newcommand{\complExternalVolume}{W^{\superscriptComplementaryEnergy V}}
\newcommand{\complExternalVolumeDensity}{\complExternalVolume_0}
\newcommand{\complExternalSurface}{W^{\superscriptComplementaryEnergy S}}
\newcommand{\complExternalSurfaceDensity}{\complExternalSurface_0}

\newcommand{\setStaticallyAdmissible}{\boldsymbol{\mathbb{S}}}

\newcommand{\penaltyWeight}{\lambda}
\newcommand{\penaltyTerm}{\pi}

\newcommand{\diffOperator}{F}
\newcommand{\stressFunction}{\boldsymbol{\Phi}}
\newcommand{\airy}{\varphi}

\newcommand{\domainRef}{\hat{\domain}}
\newcommand{\airyRef}{\hat{\airy}}
\newcommand{\cv}{\hat{\phi}}
\newcommand{\xRef}[1][]{\xi_{#1}}

\newcommand{\mapping}{\boldsymbol{T}}
\newcommand{\mappingJacobian}{\boldsymbol{J}}
\newcommand{\mappingJacobianInv}{\mappingJacobian^{-1}}
\newcommand{\mappingJacobianInvT}{\mappingJacobian^{-T}}
\newcommand{\stressRef}{\hat{\stress}}

\newcommand{\setFreeControlVars}{\mathbb{D}}

\newcommand{\nUnknowns}{\mathcal{N}}

\newcommand{\relError}[1]{\varepsilon_{L^2}[#1]}

\newcommand{\gravity}{g}
\newcommand{\rotationMatrix}{\mathcal{R}}

\begin{frontmatter}



\title{A Spline-Based Stress Function Approach for the Principle of Minimum Complementary Energy}


\author[ILSB]{Fabian Key\corref{cor}} 
\ead{fabian.key@tuwien.ac.at}
\cortext[cor]{Corresponding author}

\author[ILSB]{Lukas Freinberger}

\affiliation[ILSB]{organization={Institute of Lightweight Design and Structural Biomechanics (ILSB), TU Wien},
            addressline={Karlsplatz 13}, 
            city={Vienna},
            postcode={A-1040}, 
            country={Austria}}

\begin{abstract}
    In computational engineering, ensuring the integrity and safety of structures in fields such as aerospace and civil engineering relies on accurate stress prediction.
    However, analytical methods are limited to simple test cases, and displacement-based \glspl{fem}, while commonly used, require a large number of unknowns to achieve high accuracy; stress-based numerical methods have so far failed to provide a simple and effective alternative.  
    This work aims to develop a novel numerical approach that overcomes these limitations by enabling accurate stress prediction with improved flexibility for complex geometries and boundary conditions and fewer \glspl{dof}.  
    The proposed method is based on a spline-based stress function formulation for the principle of minimum complementary energy, which we apply to plane, linear elastostatics.
    The method is first validated against analytical solutions and then tested on two test cases challenging for current state-of-the-art numerical schemes---a bi-layer cantilever with anisotropic material behavior and a cantilever with a non-prismatic, parabolic-shaped beam geometry.
    Results demonstrate that our approach, unlike analytical methods, can be easily applied to general geometries and boundary conditions, and achieves stress accuracy comparable to that reported in the literature for displacement-based \glspl{fem}, while requiring significantly fewer \glspl{dof}.
    This novel spline-based stress function approach thus provides an efficient and flexible tool for accurate stress prediction, with promising applications in structural analysis and numerical design.  
\end{abstract}




\end{frontmatter}


\glsresetall

\section{Introduction}
\label{sec:introduction}
Accurate stress prediction is essential to computational engineering, providing the foundation for the reliable design and safety assessment of structures in aerospace, civil, and mechanical engineering applications.
Many of these applications require solving boundary value problems in elasticity, where the goal is to determine the internal state of a solid, typically described by the distributions of displacements, strains, and stresses, in response to external loads and boundary conditions.
\par
To address these problems, a variety of analytical and numerical methods have been developed in the past decades. To set the stage for our proposed method, we begin by embedding it within this broader context of commonly used solution techniques in solid mechanics. These techniques generally fall into two broad categories: (1) those based on the differential field equations of elasticity theory (see~\cite[Ch. 5]{Sadd2021}), and (2) those derived from variational principles (see~\cite{Reddy2017}).
\bigskip\par
The first approach begins with the fundamental elasticity relations, also referred to as \textit{field equations}: equilibrium equations, compatibility conditions, and the elastic constitutive law. Eliminating certain field variables yields different governing equations, depending on the choice of primary unknowns. 
For example, eliminating strains and stresses, given strain\penalty0-\hskip0pt displacement relations and the constitutive law, yields \textit{Navier's equations}, a set of second-order partial differential equations, for the displacement field to ensure equilibrium. This system is referred to as a \textit{displacement formulation} and provides three equations for the three unknown displacement components.  It is particularly well-suited when displacement boundary conditions are prescribed over the entire boundary of the domain.
Conversely, eliminating strains and displacements---again using the strain-displacement relations and the constitutive law---results in the \textit{Beltrami-Michell equations}, which govern the stress field and ensure both compatibility and equilibrium. They provide six independent equations for the six stress components and constitute a \textit{stress formulation}. Here, only traction boundary conditions can be imposed directly.
To reduce the complexity of the stress formulation, the concept of \textit{stress functions} is often employed.
In this concept, the stress components are related to the derivatives of the  (generally tensor-valued) stress function such that the equilibrium equations are automatically satisfied. As a result, the original set of six equations can be replaced by a smaller set of equations for the stress function components. The number and form of these functions are problem-dependent; for example, in the two-dimensional case, the so-called \textit{Airy stress function}~\cite{Airy1863} expresses the stress components in terms of a single scalar function.
\par
Based on these differential equations, analytical methods have historically played a central role in understanding elastic behavior by offering exact closed-form solutions. However, they are generally only applicable to problems with simple geometries, homogeneous material properties, and idealized boundary conditions.
Among these methods, we highlight the power series approach~\cite{Neou1957}, which is particularly well-suited for regular domains and will later be used to validate our approach. This method offers a systematic way to construct a polynomial approximation of the Airy stress function by applying the biharmonic compatibility equation---which results from substituting the stress function into the Beltrami-Michell equations---together with the boundary conditions. This process reduces the original doubly infinite power series to a finite polynomial form with a specific number of coefficients to be determined for the given problem.
\bigskip\par
In contrast to formulations derived from the field equations, \textit{variational methods} provide an alternative framework by recasting elasticity problems into problems of minimizing or finding stationary values of functionals, including those based on virtual work, energy minimization, or mixed formulations such as the de-Veubeke-Hu-Washizu principle~\cite{deVeubeke1965,Zienkiewicz2001,Hu1954,Washizu1955}.
While a variety of variational principles exist, we focus here on two fundamental approaches particularly relevant for the numerical treatment of solids: (1) the \textit{principle of virtual displacements}, associated with the \textit{principle of minimum potential energy}, which forms the basis of displacement-based methods; and (2) the \textit{principle of virtual forces}, together with the \textit{principle of minimum complementary energy}, which underpins stress-based formulations.
Other mixed formulations involving multiple independent fields---such as the \textit{Hellinger-Reissner} principle~\cite[p.~654--655]{Hellinger1914},~\cite{Reissner1950,Reissner1965}, which introduces displacements and stresses as independent variables---are especially relevant for plate and shell models, but are not discussed further here.
\par
These principles form the basis for a variety of numerical methods that deliver approximate solutions by applying the minimization or stationarity condition to determine the unknown coefficients in a chosen ansatz, often expressed using a finite set of basis functions.
The most widely used approach is the \textit{displacement-based \gls{fem}}, which is based on the principle of virtual displacements (see, e.g.,~\cite[Ch.~2]{Hughes2000}), closely related to the principle of minimum potential energy. The \gls{fem} approximates the displacement field using piecewise polynomial basis functions over a mesh of finite elements, resulting in a sparse system of algebraic equations. 
Due to its robustness, flexibility, and ability to handle complex geometries and boundary conditions, \gls{fem} has become the standard computational tool in structural mechanics and many other engineering disciplines.
However, stresses are obtained only indirectly by differentiation of the displacement solution. Since standard \gls{fem} employs basis functions with only $C^0$ inter-element continuity, the resulting stress field is generally discontinuous across element boundaries or requires post-processing with stress recovery techniques. As a result, achieving accurate stress predictions often necessitates fine meshes and, consequently, a large number of \glspl{dof}. 
An approach that provides higher inter-element continuity is \textit{\gls{iga}}~\cite{Hughes2005,Cottrell2009}. It uses spline-based basis functions, such as B-Splines or \gls{nurbs} with increased smoothness. This enhanced continuity improves the quality of the displacement field, which in turn leads to more accurate stress approximations. Still, \gls{iga} is predominantly applied as a displacement-based method in solid mechanics.
\bigskip
\par
While displacement-based methods only allow to compute stresses indirectly, stress-based methods directly approximate the stress field. These methods are based on the principle of virtual forces or the principle of minimum complementary energy.
To this end, \textit{stress-based \glspl{fem}} have been developed.
Early works introduced stress elements with assumed stress functions that satisfy the equilibrium conditions in their interior~\cite{Pian1964,deVeubeke1965,deVeubeke1967}. To determine the unknown coefficients, the algebraic equations were derived from the principle of minimum complementary energy, using the necessary condition of stationarity. Subsequent work extended this framework to applications with material nonlinearity, such as the incremental complementary energy approach proposed in~\cite{Rybicki1970}, which used a stress function with 36 unknowns per rectangular element. Similarly, extensions to elastoplastic analysis were presented in~\cite{Gallagher1971}. For a more comprehensive review of stress-function formulations using the complementary energy principle within \gls{fem}, we refer the reader to the overview in~\cite[p.~117]{Gallagher1993}
\par
Later developments in stress-based \gls{fem} focused on refining element formulations and improving the treatment of equilibrium and boundary conditions. 
For example, penalty formulations have been studied to incorporate inter-element constraints on stress-function elements, but choosing the right penalty factor is critical for accuracy and remains a nontrivial task~\cite{Carey1982}.
Additionally, rectangular elements with 24 \glspl{dof} and triangular elements with 18 \glspl{dof} were presented, enforcing traction boundary conditions via Lagrange multipliers~\cite{Vallabhan1982}. 
While effective, the use of Lagrange multipliers leads to a significant increase in the number of equations. 
To address the complications of prescribing traction boundary conditions, a class of element families was developed in which such conditions are satisfied directly: the so-called \glspl{bte} and \glspl{hre}, with single elements ranging from 16 to 24 \glspl{dof}~\cite{Sarigul1989,Gallagher1993}.
Furthermore, an axisymmetric formulation for the analysis of circular cylindrical shells under linear deformation was presented~\cite{Bertoti1993}. In this approach, the three-dimensional equilibrium equations and traction boundary conditions are satisfied a priori by a tensor-valued stress function, rendering Lagrange multipliers or penalty terms unnecessary.
Recent years have seen a continued interest in the development and refinement of stress-based \glspl{fem}. Building on the work in~\cite{Pian1964}, a \gls{hsf} element method for planar problems has been developed, introducing 8- and 12-node quadrilateral elements with 16 to 24 \glspl{dof}~\cite{Cen2011}.
Additionally, a strategy for modeling singular stress fields in unilateral materials that is capable of capturing stress singularities independently of the mesh resolution has been presented~\cite{Montanino2022,Montanino2024}.
Further elements have been designed for different applications such as thermoelastic triangular elements~\cite{Pour2023} or beam elements with five and six \glspl{dof} per element and using Lagrange multipliers~\cite{Wieckowski2021}.
\par
Beyond the development of new elements, a complementary energy variational approach has been presented, which approximates plane elastic problems, i.e., the behavior of a 2D continuum body, through truss structures and supports stress concentrations or singularities~\cite{Fraternali2001,Fraternali2002}.
Furthermore, a machine learning approach for linear elasticity---referred to as \gls{dcem}---has also been presented, in which the stress function is parameterized using three separate neural networks~\cite{Wang2024}. Nonetheless, this method entails considerable training effort, especially to accurately predict stresses and generalize to different loading scenarios.
\bigskip\par
Despite demonstrating improved accuracy in stress prediction compared to displacement-based methods, stress-based formulations based on complementary energy principles faced notable limitations. Many of the proposed elements were tailored to specific problem classes, limiting their general applicability. 
Additionally, the requirement to satisfy equilibrium and boundary conditions often resulted in a substantial number of equations in the final system, for example due to the high number of \glspl{dof} per element.
\bigskip
\par
Lastly, we would like to mention an approach that does not fully align with the previous distinction between differential and variational methods, i.e., the application of the \gls{fem} directly to the Beltrami–Michell equations~\cite{Sky2024a,Sky2024b}. Note that in this work, the equations require additional stabilization strategies to handle mixed boundary conditions effectively.
\bigskip
\par
Despite the variety of methods available for stress prediction in elastic problems, key challenges remain.
Analytical approaches following formulations based on the differential field equations are restricted to simple geometries and loading conditions, limiting their relevance for practical engineering applications.
Following variational principles, displacement-based methods---including standard \gls{fem} and \gls{iga}---approximate the displacement field and compute stresses via post-processing. As a result, accurate stress prediction often requires fine meshes or additional recovery procedures, increasing computational effort.
Alternatively, stress-based methods, which approximate the stress field directly, offer improved stress accuracy. However, they typically rely on specialized elements with many \glspl{dof} or involve complicated enforcement of equilibrium and boundary conditions. This lack of generality and efficiency has so far hindered their practicability and broader adoption.
\par
These observations indicate there is still need for a more systematic and computationally efficient approach to direct stress approximation.
Such an approach should maintain accuracy and handle general geometries and boundary conditions, both without excessive discretization effort.
\bigskip\par
This work aims to close this gap by introducing a novel variational method that directly approximates the stress field while ensuring equilibrium. Notably, the approach is broadly applicable to diverse geometries and loading conditions, as well as efficient in terms of the number of unknowns.
The proposed approach is based on the principle of minimum complementary energy and employs a spline-based representation of the Airy stress function, yielding a direct stress-based methodology for solving linear and plane elasticity problems. By using a stress function, equilibrium is satisfied by construction, as in earlier complementary energy methods.
Building on these foundations, the method introduces two key enhancements:
\begin{enumerate}
    \item The properties of B-splines enable the direct incorporation of traction boundary conditions into the ansatz and promote smooth stress fields
    \item A general geometric mapping allows the spline formulation to be applied to complex geometries through transformation from a reference domain
\end{enumerate}
Collectively, these features result in an accurate and efficient method for stress prediction, offering high flexibility with respect to geometry and boundary conditions.

\section{Methodology: Variational Principle and Spline-Based Approximation}
\label{sec:materialAndMethods}
We begin with a presentation of the principle of minimum complementary energy, which forms the theoretical foundation of our approach. Subsequently, we introduce the novel spline-based stress function formulation, specifically expressing the Airy stress function using B-splines. By combining the complementary energy principle with the spline-based representation, we derive an algebraic system of equations for the spline control variables that define the stress function, from which the stress field is then obtained directly through differentiation.
\subsection{The Principle of Minimum Complementary Energy}
\label{subsec:principleMinCompEnergy}
The starting point for our variational approach is the \textit{principle of minimum complementary energy}~\cite{Engesser1889,Westergaard1942,Reddy2017}, which formulates a minimization problem in terms of the \textit{stress field} $\stressTensorComponent{ij}$.  
We consider the static case and an elastic body $\domain$ with its boundary denoted by $\boundary$, where surface traction $\tractionPrescribed_i$ and displacement $\displacementPrescribed_i$ are prescribed on the portions $\boundaryTraction$ and $\boundaryDisp$, respectively. 
Furthermore, we assume small deformations and infinitesimal strains, and that the relevant energy potentials exist.
The principle requires to consider only stress fields $\stressTensorComponent{ij}$ that are \textit{statically admissible}, and we denote the set of all statically admissible stress fields by $\setStaticallyAdmissible$. In particular, this means that they are in \textit{equilibrium} with a given body force density $\bodyForceDensity_{i}$ in $\domain$ and satisfy the \textit{traction boundary conditions} on $\boundaryTraction$:
\begin{align}
    \stressTensorComponent{ij,j}
    + \bodyForceDensity_{i}
    &=
    0
    \quad\text{in } \domain,
    \label{eq:equilibrium}
    \\
    \stressTensorComponent{ij}\normal_{j}
    &=
    \tractionPrescribed_i
    \quad\text{on } \boundaryTraction,
    \label{eq:tractionBC}
\end{align}
\added{where $\normal_{j}$ denotes the outward-pointing unit normal vector.}
Then, it can be shown for conservative systems that the total complementary energy $\totalComplEnergy$ takes a stationary value for the equilibrium configuration of the body:
\begin{equation}
    \delta\totalComplEnergy
    = 0.
\end{equation}
With the additional condition
\begin{equation}
    \delta^2\totalComplEnergy > 0,
    \label{eq:secondOrderVariationPositive}
\end{equation}
the principle of minimum complementary energy can be formulated as given in 
\cite[Theorem 15b]{Hoff1956}:
\begin{quote}
    The total complementary potential is a minimum with respect to variations in stress when the system is in its true state of equilibrium.
\end{quote}
In this sense, we refer to the total complementary energy as a functional of the stresses $\totalComplEnergy\left[\stressTensor\right]$ and state the elastic problem as
\begin{equation}
    \min_{\stressTensor\in\setStaticallyAdmissible}
    \lcb
        \totalComplEnergy\lb\stressTensor\rb
    \rcb.
    \label{eq:structuralAnalysisProblem}
\end{equation}
\par
The total complementary energy $\totalComplEnergy$ can be divided into \textit{internal} and \textit{external} contributions $\complStrainEnergy$ and $\complExternal$, respectively:
\begin{equation}
    \totalComplEnergy\left[\stressTensor\right]
    =
    \complStrainEnergy\left[\stressTensor\right]
    +
    \complExternal\left[\stressTensor\right].
    \label{eq:totalComplEnergy}
\end{equation}
For the linear elastic case, we use the generalized Hooke’s law such that the internal complementary energy $ \complStrainEnergy$ is expressed as
\begin{equation}
    \complStrainEnergy
    =
    \frac{1}{2}
    \intDomain{%
        \complianceTensorComponent{ijkl}
        \stressTensorComponent{ij}
        \stressTensorComponent{kl}
    },
    \label{eq:complStrainEnergy}
\end{equation}
where $\complianceTensorComponent{ijkl}$ denotes the \textit{compliance tensor}.
Note that \Cref{eq:secondOrderVariationPositive} follows if the compliance tensor $\complianceTensor$ is positive definite.
For completeness, we also present the \added{relevant} expression\deleted{s} for the external complementary energy \added{contribution, denoted by $\complExternalSurface$, which depends on the unknown stress field $\stressTensorComponent{ij}$}:
\begin{equation}
    \complExternal
    =
    \complExternalSurface
    =
    - \intBoundaryDisp{%
        \displacementPrescribed_i
        \underbrace{\stressTensorComponent{ij}\normal_j}_{\traction_i}
    }.
    \label{eq:complementaryEnergyExternal}
\end{equation}

\subsection{Spline-Based Stress Function Approach}
\label{subsec:splineBasedStressFunction}
Building on the principle of minimum complementary energy outlined above, we now present the spline-based stress function approach.
Since this principle requires the stress field to be statically admissible, both the equilibrium equations and the traction boundary conditions given in \Cref{eq:equilibrium,eq:tractionBC} must be satisfied.
As discussed in the introduction, the use of stress functions to enforce equilibrium is well established. Here, we extend this idea by proposing a spline-based formulation of the stress function to additionally satisfy the traction boundary conditions by construction.
\subsubsection{Stress Functions and the Airy Stress Function}
\label{subsubsec:airyStressFunction}
The concept of stress functions aims to satisfy the equilibrium conditions by providing a representation of the stress field in the following form~\cite[Ch. 13.6]{Sadd2021}:
\begin{equation}
    \stressTensorComponent{ij} = \diffOperator_{ij}\left\{\stressFunction\right\},
\end{equation}
with $\diffOperator_{ij}$ and $\stressFunction$ being a differential operator and the tensor-valued stress function, respectively.
If these elements are chosen in a way that the equilibrium equations from \Cref{eq:equilibrium} are always satisfied, it is given in a \textit{self-equilibrated} form.
\bigskip\par
For plane strain and plane stress problems, the \textit{Airy stress function}~\cite{Airy1863} is a special scalar-valued stress function that we denote by $\airy=\airy(x,y)$. At this stage, no assumptions are made regarding the form of $\airy$, which is treated as arbitrary. If we assume that the body forces can be derived from a \textit{potential} $V$ as
\begin{equation}
    \bodyForceDensity_x = -\frac{\partial V}{\partial x},\quad \bodyForceDensity_y = -\frac{\partial V}{\partial y},
    \label{eq:bodyForceDensity}
\end{equation}
the components of the stress tensor $\stressTensor$ are defined through second-order derivatives of $\airy$:
\begin{equation}
    \stress_{x}  = \frac{\partial^2\airy}{\partial y^2} + V,\quad 
    \stress_{y}  = \frac{\partial^2\airy}{\partial x^2} + V,\quad
    \stress_{xy} = -\frac{\partial^2\airy}{\partial x \partial y}.
    \label{eq:stressAiry}
\end{equation}
It can be easily checked that this representation identically satisfies the equilibrium equations.
What remains is to determine a specific form of $\airy$ such that the corresponding stress field $\stressTensorComponent{ij}$ also satisfies the traction boundary conditions given in \Cref{eq:tractionBC}, thereby completing the static admissibility. In the following, we present how the proposed spline-based ansatz can be used to meet this requirement.
\subsubsection{B-Spline Representation of the Airy Stress Function}
\label{subsubsec:bSplineRepresentation}
\begin{figure}
    \centering
    \includegraphics{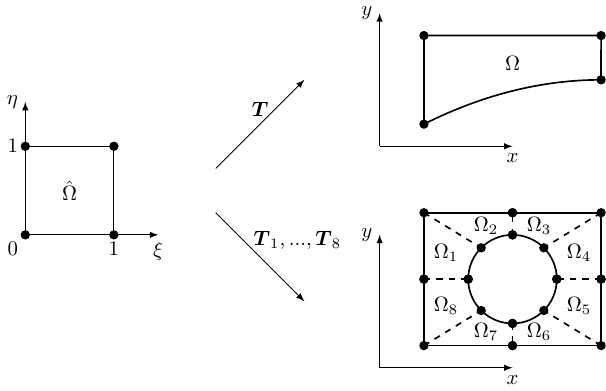}
    \caption{Geometric mappings from the parametric domain $\domainRef$. Top: mapping to a quadrilateral domain $\domain$ with curved boundary using a single mapping $\mapping$. Bottom: Mapping to a plate with a circular hole, constructed from eight patches $\Omega_1$ to $\Omega_8$ with individual mappings $\mapping_1$ to $\mapping_8$. Interfaces are indicated by dashed lines and coupled through traction equilibrium.}
    \label{fig:mappingParametricPhysical}
\end{figure}

To enable the spline-based representation of the Airy stress function $\airy(x,y)$ on general, complex geometries, we begin by introducing a \textit{geometric mapping} $\mapping$ from a simple reference domain $\domainRef$, defined in parametric coordinates $(\xi,\eta)$, to the physical domain $\domain$ of the elastic body, i.e., $\mapping: (\xi,\eta)\mapsto(x,y)$. 
An illustration of this geometric mapping is shown in the top part of \Cref{fig:mappingParametricPhysical}.
\par
Complex domains, including those with holes or multiple subregions, can be handled via multipatch formulations, where each patch has its own parametric mapping and patches are coupled by enforcing equilibrium of interface traction. Such an example is shown in the bottom part of \Cref{fig:mappingParametricPhysical}.
\par
The geometric mapping provides the foundation for expressing the stress function using B-spline basis functions defined over the parametric space. Note that this mapping is independent of the chosen ansatz for the stress function and does not have to be spline-based, although a spline-based mapping can be employed if desired. In this respect, the capabilities of the method are comparable to those of \gls{iga}, which also uses spline-based representations for the geometry and, in addition, for the solution fields. Consequently, the proposed approach is general and versatile with respect to the domains that can be modeled by corresponding geometric mappings.
\bigskip\par
Having established the geometric mapping between the parametric and physical domains, we define the Airy stress function $\airyRef(\xi,\eta)$  on the parametric domain such that it corresponds exactly to the physical stress function $\airy(x,y)$ through the relation $\airyRef(\xi,\eta)=\airy(x(\xi,\eta),y(\xi,\eta))$.
This definition enables us to represent $\airyRef$ using B-spline basis functions naturally defined over the simple reference domain.
The B-spline representation of $\airyRef$ is then given as follows (see~\cite[Equation (2.17)]{Cottrell2009}, for example):
\begin{equation}
    \airyRef(\xi,\eta) = \sum_{i=1}^{n} \sum_{j=1}^{m} N_{i,p}\lp\xi\rp M_{j,q}\lp\eta\rp \cv_{i,j},
\end{equation}
where $N_{i,p}(\xi)$ and $M_{j,q}(\eta)$ are univariate B-spline \textit{basis functions} of \textit{polynomial degrees} $p$ and $q$, respectively. These basis functions form a smooth, flexible basis over the reference domain $\domainRef=[0,1]\times[0,1]$.
The coefficients $\cv_{i,j}$, called \textit{control variables}, parameterize the stress function in terms of this basis. They represent the \glspl{dof} that will be determined through the variational formulation, effectively defining the shape of the Airy stress function and thereby the stress field.
The B-spline basis functions are constructed from \textit{knot vectors}, i.e., non-decreasing sequences of coordinates $\xi$ and $\eta$, given as $\Xi=[\xi_1,\dots,\xi_{n+p+1}]$ and $\mathcal{H}=[\eta_1,\dots,\eta_{m+q+1}]$, respectively. These knot vectors divide the reference domain in so-called \textit{knot spans} or \textit{elements} and, together with the control variables, represent the discretization of $\airyRef$.
By combining the geometric mapping from the parametric domain to the physical domain with this spline representation, the Airy stress function can be efficiently approximated on complex geometries with a controllable level of smoothness and accuracy.
\subsection{Formulation and Solution of the Elastic Problem}
\label{subsec:formulationAndSolutionOfTheElasticProblem}
Based on the principle of minimum complementary energy and the spline-based representation of the Airy stress function introduced above, this section explains how to formulate and solve the resulting discrete elastic problem.
First, we describe how to directly incorporate traction boundary conditions into the B-spline ansatz for the stress function, thereby ensuring the static admissibility of the solution.
Then, we apply the complementary energy principle to solve for the unknown control variables, allowing us to fully define the stress function and consequently predict the stress field.
\subsubsection{Traction Boundary Conditions}
\label{subsubsec:tractionBoundaryConditions}
To incorporate the physical traction boundary conditions into the B-spline ansatz for the Airy stress function $\airyRef$, the prescribed traction must first be transformed from the physical coordinates $(x,y)$ to the parametric coordinates $(\xi,\eta)$. 
Because the Airy stress function defines the physical stress components through its second-order derivatives, the traction boundary conditions directly translate into conditions on the second-order derivatives of the stress function $\airy$ in physical coordinates. 
To transfer them to $\airyRef$, i.e., express them in terms of the parametric coordinates, we apply the chain rule and relate the second-order derivatives in physical space to a combination of first- and second-order derivatives in the parametric space, as well as derivatives of the geometric mapping $\mapping$.
To this end, we define $\mappingJacobian$ as the Jacobian of the geometric mapping $\mapping$ and its inverse $\mappingJacobianInv$ as follows:
\begin{equation}
    \mappingJacobian(\xi,\eta) 
    = \frac{\partial \mapping(\xi,\eta)}{\partial(\xi,\eta)}
    =
    \lb
    \begin{array}{cc}
         \frac{\partial x}{\partial \xi}& \frac{\partial x}{\partial \eta}  \\
         \frac{\partial y}{\partial \xi}& \frac{\partial y}{\partial \eta}
    \end{array}
    \rb
    \quad
    \text{and}
    \quad
    \mappingJacobianInv(\xi,\eta) 
    =
    \lb
    \begin{array}{cc}
         \frac{\partial \xi}{\partial x}& \frac{\partial \xi}{\partial y}  \\
         \frac{\partial \eta}{\partial x}& \frac{\partial \eta}{\partial y}
    \end{array}
    \rb.
\end{equation}
Then, the following holds for the second-order derivatives of $\airy$ in physical space:
\begin{equation}\label{eq:secondOrderDerivatives}
\begin{split}
    \left[
    \begin{array}{cc}
         \frac{\partial^2 \airy}{\partial x^2}&  \frac{\partial^2 \airy}{\partial x\partial y}\\
         \frac{\partial^2 \airy}{\partial x\partial y} & \frac{\partial^2 \airy}{\partial y^2}
    \end{array}
    \right]
    &=
    \mappingJacobianInvT
    \lb
    \begin{array}{cc}
         \frac{\partial^2 \airyRef}{\partial \xi^2}&  \frac{\partial^2 \airyRef}{\partial \xi\partial \eta}\\
         \frac{\partial^2 \airyRef}{\partial \xi\partial \eta} & \frac{\partial^2 \airyRef}{\partial \eta^2}
    \end{array}
    \rb
    \mappingJacobianInv
    \\
    &+
    \lb
    \begin{array}{cc}
         \frac{\partial^2\xi}{\partial x^2}\frac{\partial\airyRef}{\partial\xi} + \frac{\partial^2\eta}{\partial x^2} \frac{\partial\airyRef}{\partial\eta} & \frac{\partial^2\xi}{\partial x \partial y}\frac{\partial\airyRef}{\partial\xi} + \frac{\partial^2\eta}{\partial x \partial y} \frac{\partial\airyRef}{\partial\eta}\\
         \frac{\partial^2\xi}{\partial x \partial y}\frac{\partial\airyRef}{\partial\xi} + \frac{\partial^2\eta}{\partial x \partial y} \frac{\partial\airyRef}{\partial\eta} & \frac{\partial^2\xi}{\partial y^2}\frac{\partial\airyRef}{\partial\xi} + \frac{\partial^2\eta}{\partial y^2} \frac{\partial\airyRef}{\partial\eta} 
    \end{array}
    \rb
\end{split}
\end{equation}
As a result, since the geometric mapping and all its derivatives are known and fixed by construction, the traction boundary conditions on the physical stress components---virtually given on the left-hand side of \Cref{eq:secondOrderDerivatives}---translate directly into constraints on the derivatives of $\airyRef$ and, consequently, on its B-spline representation.
\bigskip\par
To build these constraints on $\airyRef$ into its B-spline ansatz, we derive conditions on the control variables $\cv_{i,j}$. To this end, we make use of the property that the derivatives of a B-spline are also B-splines, albeit of lower degrees, and the control variables of the derivative B-splines are related to the control variables of the original B-spline through linear combinations.
For example, consider the first-order derivative of $\airyRef$ with respect to the parametric coordinate $\xi$.
Due to the differentiability properties of B-splines explained above, this derivative can be expressed as follows:
\begin{equation}
    \frac{\partial \airyRef}{\partial \xi} 
    =
    \sum_{i=1}^{n-1}
    \sum_{j=1}^{m} 
    N_{i,p-1}\lp\xi\rp M_{j,q}\lp\eta\rp \cv^{\xi}_{i,j},
\end{equation}
where $N_{i,p-1}(\xi)$ are B-spline basis functions of degree $p-1$ constructed from a modified knot $\Xi^\xi$, obtained by removing the first and last entries from the original knot vector $\Xi$. The control variables $\cv^{\xi}_{i,j}$ associated with the derivative B-spline $\frac{\partial \airyRef}{\partial \xi}$ are linear combinations of the original control variables $\cv_{i,j}$ of $\airyRef$.
More specifically, the derivative control variables are given by
\begin{equation}
    \cv^{\xi}_{i,j} = p \frac{\cv_{i+1,j}-\cv_{i,j}}{\xi_{i+p+1}-\xi_{i+1}},
    \label{eq:controlVariableDerivative}
\end{equation}
where $\xi_{i+p+1}$ and $\xi_{i+1}$ are entries of the original knot vector $\Xi$.
Now, suppose we want to enforce a condition on the derivative $\frac{\partial \airyRef}{\partial \xi}$ by prescribing the values of $ \cv^{\xi}_{i,j}$. Since the right-hand side of \Cref{eq:controlVariableDerivative} depends linearly on the original control variables $\cv_{i+1,j}$ and $\cv_{i,j}$, such a condition can be directly translated into a linear constraint on these original control variables.
\par
\begin{figure}
    \centering
    \includegraphics[width=\textwidth]{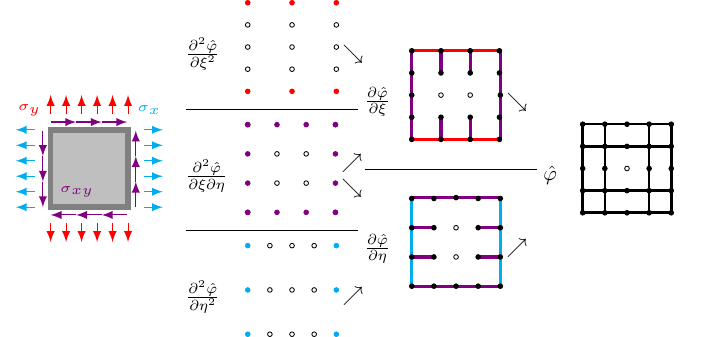}
    \caption{Incorporation of traction boundary conditions for the special case $\mapping=\rm id$: prescribed stress components (left) and control variables for the spline-based stress function $\airyRef$ and its derivatives (right). The prescribed stress components determine the values of the marked control variables (filled circles) of the second-order derivative splines. These conditions are propagated to $\airyRef$, imposing constraints on a subset of the control variables $\cv_{i,j}$. These constraints are indicated by lines between involved control variables. Note that, in this example, only the middle control variable of $\airyRef$ remains unconstrained.}
    \label{fig:boundaryConditions}
\end{figure}

This reasoning can be generalized to higher-order derivatives as well. Therefore, any condition involving first- and second-order derivatives of $\airyRef$---such as those arising from traction boundary conditions---can be systematically propagated to a corresponding linear constraint on a subset of control variables $\cv_{i,j}$.
In particular, assembling all such conditions yields a linear system of equations that can be solved for this subset of control variables that influence the boundary behavior.
After solving this system, the remaining control variables---those not involved in satisfying the traction boundary conditions---remain as the unknowns to be determined through the complementary energy principle.
This process of incorporating the traction boundary conditions is illustrated in \Cref{fig:boundaryConditions}, where, for simplicity, the special case of the identity mapping $\mapping=\rm id$ is considered. 
\bigskip\par
So, the procedure allows us to enforce certain boundary conditions in this strong sense.
However, we note that, depending on the specific form of the prescribed traction, the B-spline ansatz may not be able to satisfy all conditions exactly, but only in a weak sense.
Please note that, because the B-spline basis is non-interpolatory, the boundary values cannot be directly imposed at control variables as in standard finite elements. Instead, the boundary conditions are enforced, e.g., in a least-squares sense, similar to common strategies in isogeometric analysis~\cite[p.~85]{Cottrell2009}.
\par
To address this, we introduce quantities that allow for the weak enforcement of a locally varying traction, a prescribed resultant force, or more general conditions.
First, we define a measure $\mathfrak{D}(\cv_{i,j})$ for the squared point-wise differences between the traction resulting from our B-spline ansatz, $\traction_i(\cv_{i,j})$, and the prescribed traction, $\tractionPrescribed_i$, which may vary arbitrarily along the boundary.
By minimizing this measure, we enable the B-spline to approximate the boundary conditions as accurately as possible:
\begin{equation}
    \mathfrak{D}\lp\cv_{i,j}\rp = \int_{\boundaryTraction} \left|\traction_i\lp\cv_{i,j}\rp-\tractionPrescribed_i\right|^2\,d\boundary \rightarrow \min
    \label{eq:tractionDifferenceIntegral}
\end{equation}
Similarly, we define a squared residual $\mathfrak{R}(\cv_{i,j})$ that quantifies the difference between the resultant force computed from the B-spline representation and a prescribed total force $F_i$, which we then aim to minimize:
\begin{equation}
    \mathfrak{R}\lp\cv_{i,j}\rp = \lp\int_{\boundaryTraction} \traction_i\lp\cv_{i,j}\rp\,d\boundary - F_i\rp^2 \rightarrow \min
    \label{eq:resultantForceIntegral}
\end{equation}
Note that in addition to the pointwise or resultant-based traction conditions discussed above, more general constraints, such as the prescription of moments, can likewise be enforced weakly through additional terms $\mathfrak{C}(\cv_{i,j})$.
To determine the control variables that best approximate the prescribed conditions in a weak sense, we compute the minimizer of the combined error measures $\mathfrak{D}(\cv_{i,j})$, $\mathfrak{R}(\cv_{i,j})$, and $\mathfrak{C}(\cv_{i,j})$. 
This leads to the following necessary condition for optimality:
\begin{equation}
    \frac{\partial \lp
        \mathfrak{D}\lp\cv_{i,j}\rp 
        + 
        \mathfrak{R}\lp\cv_{i,j}\rp 
        +
        \mathfrak{C}\lp\cv_{i,j}\rp 
        \rp}
    {\partial\cv_{i,j}}
    = 
    0
    \label{eq:tractionBCsOptimality}
\end{equation}
As before, solving the resulting equations determines the values of the control variables involved in the weak enforcement of the traction boundary conditions, while leaving the remaining ones unconstrained.
\par
We would like to stress that this strategy is general, as it can accommodate arbitrary traction data (see \Cref{eq:tractionDifferenceIntegral}), resultant forces (see \Cref{eq:resultantForceIntegral}), or more general constraints through the minimization procedure, while displacement conditions are incorporated naturally through the complementary energy (see \Cref{eq:complementaryEnergyExternal}). Since the formulation and solution procedure for the linear system of equations remains unchanged, standard and repeatable applicability is ensured.
\bigskip\par
In distinction to the control variable constrained in this way, we denote the set of remaining ``free'' control variables by $\setFreeControlVars$. These variables, $\cv_{i,j} \in \setFreeControlVars$, now serve as the \glspl{dof} in the subsequent minimization of the complementary energy, which we detail in the following section.
\subsubsection{Mimimization of the Complementary Energy}
\label{subsubsec:minimizationOfTheComplementaryEnergy}
In the following, we integrate the spline-based stress function into the minimization of the complementary energy to complete the solution of the elastic problem. After incorporating the traction boundary conditions, the remaining unknowns are the control variables $\cv_{i,j}\in\setFreeControlVars$ and we denote the stress field accordingly as $\stressTensor = \stressTensor(\cv_{i,j})$.
We recall that the use of a stress function, along with the enforced traction boundary conditions, guarantees static admissibility of the stress field, i.e., $\stressTensor(\cv_{i,j})\in\setStaticallyAdmissible$.  
It is also important to recall that displacement boundary conditions enter the formulation through the external contribution $\complExternal$ to the total complementary energy (see \Cref{eq:complementaryEnergyExternal}). 
Consequently, we can reformulate the elastic problem by transforming \Cref{eq:structuralAnalysisProblem} into an unconstrained minimization of the total complementary energy with respect to the remaining control variables $\cv_{i,j} \in \setFreeControlVars$:
\begin{equation}
    \min_{ \cv_{i,j}\in\setFreeControlVars}
    \lcb
        \totalComplEnergy
        \lb
            \stressTensor\lp\cv_{i,j}\rp
        \rb
    \rcb.
    \label{eq:elasticProblemDiscreteSpline}
\end{equation}
Rather than solving the minimization problem in \Cref{eq:elasticProblemDiscreteSpline} directly, we can instead employ the necessary condition for optimality:
\begin{equation}
    \frac{\partial \totalComplEnergy}{\partial\cv_{i,j}} = 0,
    \label{eq:elasticProblemNecessaryCondition}
\end{equation}
which yields a system of linear equations for the unknown control variables $\cv_{i,j}\in\setFreeControlVars$.
We note that if the traction boundary conditions are enforced weakly, the necessary conditions for optimality from \Cref{eq:tractionBCsOptimality,eq:elasticProblemNecessaryCondition} (boundary conditions and complementary energy) can also be combined into a single system.
\par
Since we consider only plane problems, we adopt the following matrix-vector notation for the stress–strain relationship, relating the vectors of stress and strain components given by $(\stressTensorComponent{xx},\stressTensorComponent{yy},\stressTensorComponent{xy})^T$ and $(\strainTensorComponent{xx},\strainTensorComponent{yy},\strainTensorComponent{xy})^T$, respectively:
\begin{equation}
    (\strainTensorComponent{xx},\strainTensorComponent{yy},\strainTensorComponent{xy})^T
    =
    \complianceMatrix
    \,
    (\stressTensorComponent{xx},\stressTensorComponent{yy},\stressTensorComponent{xy})^T,
\end{equation}
where $\complianceMatrix$ is the compliance tensor in matrix form.
For instance, in the case of plane stress, the compliance matrix takes the form
\begin{equation}
    \complianceMatrix
    =
    \frac{1}{\youngsModulus}
    \lp
    \begin{array}{ccc}
         1& -\poissonsRatio & 0\\
         -\poissonsRatio& 1 & 0\\
         0& 0& 1+\poissonsRatio
    \end{array}
    \rp,
\end{equation}
where $\youngsModulus$ and $\poissonsRatio$ denote Young's modulus and Poisson's ratio, respectively.
\bigskip\par
In summary, the proposed method to solve the elastic problem proceeds by combining the spline-based representation of the stress function with the variational formulation based on complementary energy. The overall approach can be outlined as follows:
\begin{enumerate}
    \item \textbf{Define the geometric mapping:} For a given problem geometry, construct one or multiple geometric mappings $\mapping$ or $\mapping_i$, transforming the reference domain $\domainRef$ into the physical domain $\domain$ of the elastic body.
    \item \textbf{Specify the B-spline stress function}: Define the B-spline representation of the stress function $\airyRef$ in the parametric domain, choosing the polynomial degrees $p$ and $q$, the number of control variables $n$ and $m$, and the associated knot vectors $\Xi$ and $\mathcal{H}$. 
    \item \textbf{Incorporate traction boundary conditions:} Enforce the traction boundary conditions either strongly (by directly imposing constraints) or weakly (e.g., by minimizing residuals), resulting in a system of equations for the control variables $\cv_{i,j}$. Solving this system determines the constrained control variables and identifies the remaining unconstrained ones, denoted as $\cv_{i,j} \in \setFreeControlVars$.
    \item \textbf{Minimize the total complementary energy:} Using the previously defined B-spline ansatz for the stress function, the total complementary energy $\totalComplEnergy$ from \Cref{eq:totalComplEnergy} becomes a scalar expression depending only on the unconstrained control variables $\cv_{i,j} \in \setFreeControlVars$. To determine these unknowns, minimize the total complementary energy directly as formulated in \Cref{eq:elasticProblemDiscreteSpline}, or solve the associated optimality condition in \Cref{eq:elasticProblemNecessaryCondition}. This final step fully determines the stress function $\airyRef$.
    \item \textbf{Compute the physical stress field:} Finally, the components of the stress tensor $\stressTensor$ are obtained as second-order derivatives of the Airy stress function $\airy$, by evaluating the derivatives of $\airyRef$ in the parametric domain and applying the chain rule using the derivatives of the geometric mapping.
\end{enumerate}
\bigskip\par
Before applying this approach to practical test cases, we briefly highlight its key advantages.
First, the spline-based ansatz for the stress function offers flexibility in choosing polynomial degrees, ensuring high-order continuity of the stress function and, thus, a smooth stress distribution throughout the domain. This eliminates the need for additional inter-element treatment commonly required in the existing methods, allowing for relatively low number of unknowns.
Furthermore, traction boundary conditions can be seamlessly integrated into the B-spline stress function formulation by leveraging the relationship between the control variables of a B-spline and its derivatives.
Finally, the geometric mapping allows for the representation of complex-shaped domains while preserving the formulation in the parametric space, combining flexibility with conceptual simplicity.
Together, these features position the spline-based approach as a flexible and efficient alternative for solving elasticity problems using a stress function within the principle of minimum complementary energy framework.
\section{Results and Discussion}
\label{sec:resultsAndDiscussion}
To illustrate the functionality of the proposed approach, we begin by validating our method against two test cases with analytic solutions. 
This comparison focuses on the stress distributions to demonstrate the accuracy of the spline-based formulation.
Subsequently, we consider two additional test cases in which accurate stress prediction remains challenging for existing numerical methods: (1) a bi-layer cantilever with anisotropic material behavior and (2) a parabolic-shaped cantilever as an example for non-prismatic beams, illustrating the application of the approach to problems with non-regular geometries. In these examples, we evaluate both the accuracy of the resulting stress fields and the efficiency of the method by comparing the final number of unknowns required in our formulation.
\subsection{Bar under Self-Weight}
\label{subsec:barUnderSelfWeight}
\begin{figure}
    \centering
    \includegraphics[width=0.25\textwidth]{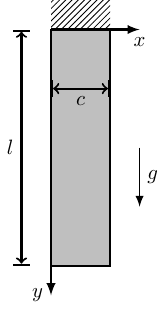}
    \caption{Bar under self-weight: setup.}
    \label{fig:barUnderSelfWeightSetup}
\end{figure}

As a first validation test case, we consider a vertical prismatic bar of length $l$ and thickness $c$, with unit out-of-plane depth and material density $\density$, subjected to self-weight. The top edge is clamped (zero displacement), while all other edges are considered traction-free. 
The setup is illustrated in \Cref{fig:barUnderSelfWeightSetup}, and all relevant quantities are listed in \Cref{tab:barUnderSelfWeightDimensionsAndLoading}. 
\begin{table}
    \centering
    \begin{tabular}{cccccc}
        \toprule
         $l$ (\si{\meter})& $c$ (\si{\meter})&$\gravity$ (\si{\meter\per\second\squared})& $\density$ (\si{\kilo\gram\per\meter\cubed})&$\youngsModulus$ (\si{\pascal}) & $\poissonsRatio$ (--)\\
         \midrule
         $2$& $0.5$ & $9.81$ & $1.0$ & $1\times10^5$& $0.3$\\
         \bottomrule
    \end{tabular}
    \caption{Bar under self-weight: dimensions, gravity, and material parameters.}
    \label{tab:barUnderSelfWeightDimensionsAndLoading}
\end{table}
We assume a uniform self-weight and, specifically, use
\begin{equation}
    V = -\density \gravity y + c_V,
\end{equation}
where the $y$-axis points downward in the direction of gravity $\gravity$, and the constant $c_V$ is set to zero without loss of generality. Consequently, the body force density from \Cref{eq:bodyForceDensity} becomes
\begin{equation}
    \bodyForceDensity_x = 0, \quad \bodyForceDensity_y  = \density \gravity.
\end{equation}
Away from the clamped top, where end effects are negligible, the analytical stress field is
\begin{equation}
    \stressTensorComponentRef{yy}\lp y\rp = \density\gravity\lp l-y\rp, \quad \stressTensorComponentRef{xx}=\stressTensorComponentRef{xy}=0,
\end{equation}
satisfying equilibrium with the uniform body force and traction-free lateral boundaries.
\par
For the spline-based solution, we define the geometric mapping $\mapping$ as
\begin{equation}
    \lp
    \begin{array}{c}
        x\\
        y
    \end{array}
    \rp
    =
    \mapping\lp\xi,\eta\rp
    =
    \lp
    \begin{array}{c}
        \xi c\\
        \lp 1-\eta\rp l 
    \end{array}
    \rp.
\end{equation}
For the stress function ansatz $\airyRef$, we choose the polynomial degrees $p=q=3$ for the B-spline basis functions. The grid of control variables $\cv_{i,j}$, the so-called control net, is defined using $n=5$ and $m=10$ control variables. Furthermore, we use uniform and open knot vectors in both dimensions. So, the total number of unknowns in our ansatz, i.e., the number of control variables, is $\nUnknowns =n\cdot m = 50$.
\par
Please note the following remark on the choice of the polynomial degrees and control variables: The B-spline degrees must be chosen such that the basis is at least twice differentiable, since the physical stress components involve parametric derivatives of the stress function up to order two. For example, cubic degrees ensure at least a linear approximation of all stress components. Additionally, the number of control variables in each parametric direction must be at least the corresponding degree plus one, ensuring the existence of at least one element. Naturally, the chosen discretization, i.e., the polynomial degrees and the number of control variables, influences both the approximation quality and the computational efficiency. While no prior knowledge of the solution is required, such knowledge may be used, to guide the choice of parameters. This is fully consistent with traditional numerical schemes such as \gls{fem} or \gls{iga}.
\bigskip\par
Next, we discuss the implementation of boundary conditions. The zero-displacement constraint at the top is enforced automatically via the complementary energy. For the remaining boundaries, which are traction-free, we minimize the residual traction (see \Cref{tab:barUnderSelfWeightWeakBoundaryConditions}) as described in \Cref{subsubsec:tractionBoundaryConditions}.
\begin{table}
    \centering
    \begin{tabular}{ccc}
        \toprule
        Left & Bottom  & Right \\
        \midrule
        $\int_{0}^{L} \left|\traction_i\right|^2\,dy$ & $\int_{0}^{c} \left|\traction_i\right|^2\,dx$& $\int_{0}^{L} \left|\traction_i\right|^2\,dy$\\
        \bottomrule
    \end{tabular}
    \caption{Bar under self-weight: quantities to be minimized for a weak enforcement of the zero-traction boundary conditions.}
    \label{tab:barUnderSelfWeightWeakBoundaryConditions}
\end{table}
\bigskip\par
\begin{figure}
    \centering
    \includegraphics[width=\textwidth]{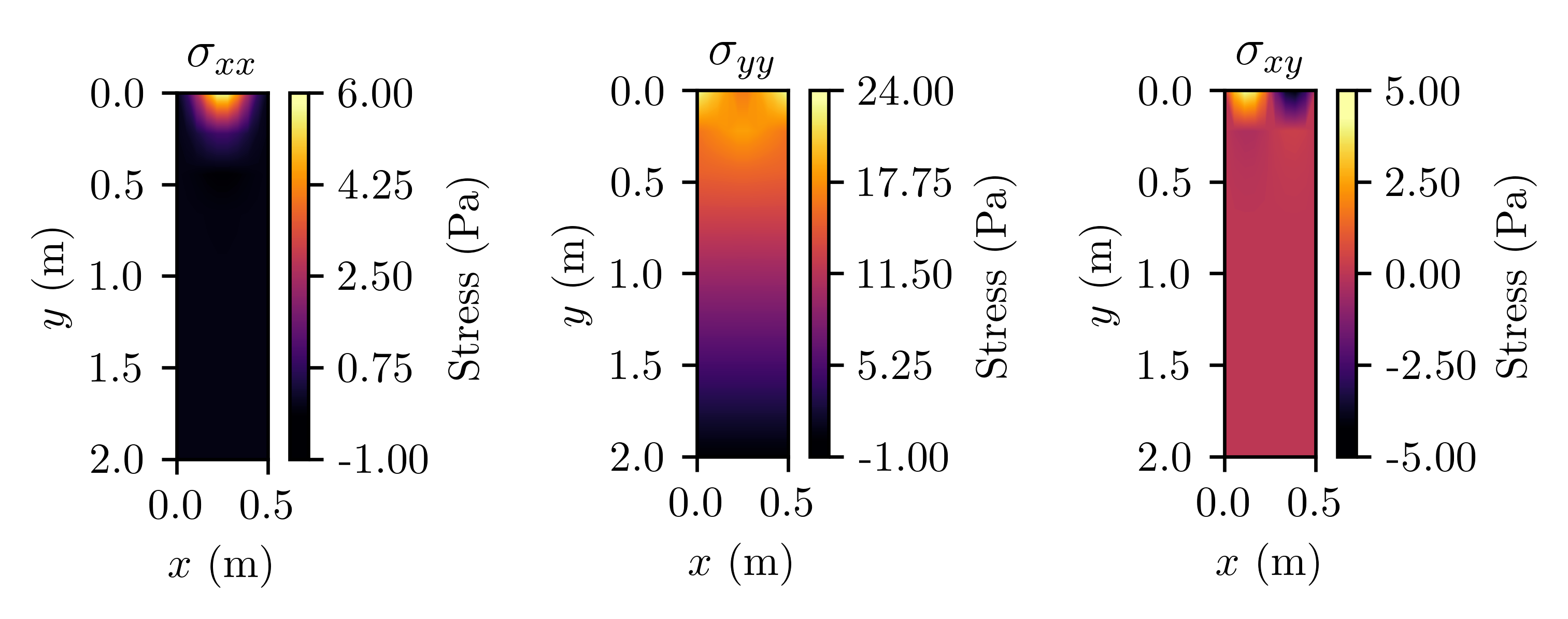}
    \caption{Bar under self-weight: stress distributions for the three stress components.}
    \label{fig:barUnderSelfWeightStress}
\end{figure}
\begin{figure}
    \centering
    \includegraphics[width=0.55\textwidth]{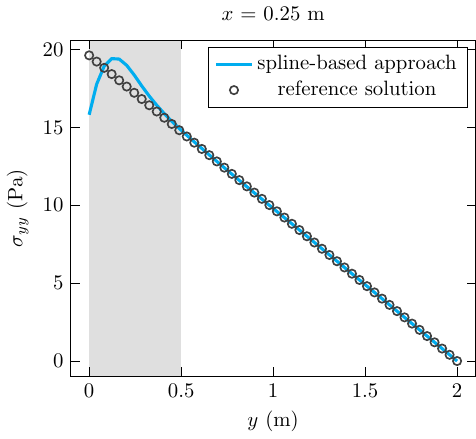}
    \caption{Bar under self-weight: comparison of the stress component $\stressTensorComponent{yy}$ across the beam’s height at mid-width ($x = 0.25~\si{\meter}$). The gray shading, equal to one cross-section dimension ($c=0.5~\si{\meter}$), marks the region near the clamped top where boundary effects are expected.}
    \label{fig:barUnderSelfWeightComparison}
\end{figure}

After enforcing the boundary conditions, the optimality condition for minimizing the total complementary energy yields a linear system, which we solve for the unconstrained control variables $\cv_{i,j}\in\setFreeControlVars$.
The stress components follow from the second-order derivatives of the stress function $\airy$ together with the potential $V$ for self-weight, as given in \Cref{eq:stressAiry}.
Figure~\ref{fig:barUnderSelfWeightStress} presents the resulting distributions of all three stress components. As expected, $\stressTensorComponent{yy}$ exhibits a linear variation with height, while $\stressTensorComponent{xx}$ and $\stressTensorComponent{xy}$ are mostly homogeneous. Deviations from this ideal behavior appear only near the clamped top, reflecting boundary effects. 
In addition, Figure~\ref{fig:barUnderSelfWeightComparison} compares the numerical and analytical solutions for $\sigma_{yy}$ across the bar height at mid-width. 
Good agreement is observed except in the region close to the clamped top, with deviations confined to a distance of about one cross-section dimension, as expected from boundary effects.
This demonstrates that the proposed method accurately captures the stress distribution and reproduces the analytical reference solution away from boundary layers.
\subsection{Bending of a Beam by Uniform Transverse Loading}
\label{subsec:bendingBeamUniformTransverseLoading}
\begin{figure}
    \centering
    \includegraphics[width=\textwidth]{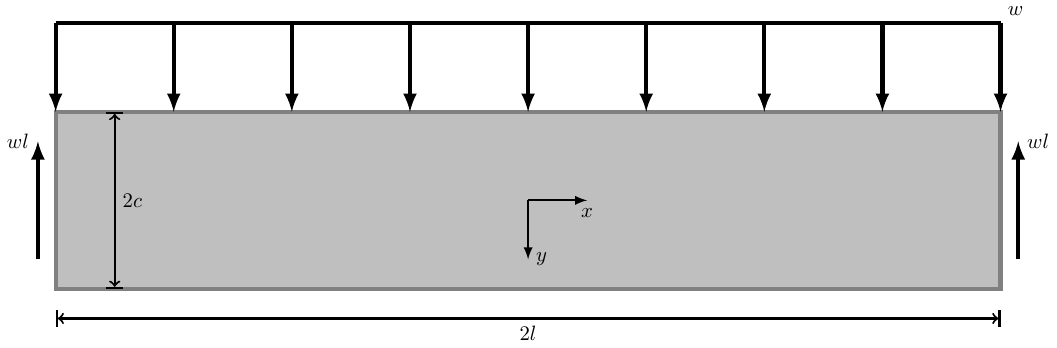}
    \caption{Bending of a beam by uniform transverse loading: setup.}
    \label{fig:bendingBeamUniformTransverseLoadingSetup}
\end{figure}

As the second validation test case, which involves more complex boundary conditions, we consider the classical problem of a rectangular beam subjected to uniform transverse loading under plane stress conditions. This example is taken from a textbook~\cite[Example~8.3]{Sadd2021}, which provides a polynomial solution following the power series approach.
We take this as a reference solution to further validate our approach by comparing the resulting stress fields.
The test case models the bending of a beam subjected to a uniform transverse load $w$ applied along its top surface.
The beam has a rectangular shape with total length $2l$ and total height $2c$, where $l$ and $c$ denote the half-length and half-height, respectively. Additionally, we assume a unit out-of-plane depth. The values for the dimensions and the loading are given in \Cref{tab:bendingBeamUniformTransverseLoadingDimensionsAndLoading}.
\begin{table}
    \centering
    \begin{tabular}{cccccc}
        \toprule
         $l$ (\si{\meter})& $c$ (\si{\meter})& $l/c$  (--)&$w$ (\si{\newton\per\meter})& $\youngsModulus$ (\si{\pascal}) & $\poissonsRatio$ (--)\\
         \midrule
         $3$& $0.25$ & $12$ & $1$ & $1\times10^5$& $0.3$\\
         \bottomrule
    \end{tabular}
    \caption{Bending of a beam by uniform transverse loading: dimensions, loading, and material parameters.}
    \label{tab:bendingBeamUniformTransverseLoadingDimensionsAndLoading}
\end{table}
The boundary conditions are given as follows.
Pointwise boundary conditions are applied on the top and bottom surfaces, while the ends of the beam are treated with simplified, statically equivalent boundary conditions. This is due to the fact that the polynomial ansatz cannot satisfy general boundary conditions exactly. To address this issue, the complex, often explicitly unknown, boundary conditions are replaced by statically equivalent polynomial conditions in accordance with the Saint-Venant principle. 
As a result, the solution is expected to be accurate primarily in regions sufficiently far from the boundaries where these simplifications are applied.
In this case, the simplified boundary conditions at the ends of the beam include zero resultant horizontal forces and bending moments, and a vertical resultant force ensuring global equilibrium. 
The complete set of boundary conditions is summarized in \Cref{tab:bendingBeamUniformTransverseLoadingBoundaryConditions} and the geometry is illustrated in \Cref{fig:bendingBeamUniformTransverseLoadingSetup}.
\begin{table}
    \centering
    \begin{tabular}{lccc}
        \toprule
                & Horizontal  & Vertical & Moment\\
        \midrule
        \rule{0pt}{20pt}
        Top     & $\stressTensorComponent{xy}=0$  & $\stressTensorComponent{y}=-w $ &-- \\
        \rule{0pt}{20pt}
        Bottom  & $\stressTensorComponent{xy}=0$   & $\int_{-l}^{l} \stressTensorComponent{y}=0 $          &--\\
        \rule{0pt}{20pt}
        Left    & $\int_{-c}^{c} \stressTensorComponent{x}\,dy=0$   & $\int_{-c}^{c} \stressTensorComponent{xy}\,dy =+ wl$& $\int_{-c}^{c} \stressTensorComponent{x}y\,dy=0$\\
        \rule{0pt}{20pt}
        Right   & $\int_{-c}^{c} \stressTensorComponent{x}\,dy=0$   & $\int_{-c}^{c} \stressTensorComponent{xy}\,dy=-wl $&$\int_{-c}^{c} \stressTensorComponent{x}y\,dy=0$\\
        \bottomrule
    \end{tabular}
    \caption{Bending of a beam by uniform transverse loading: boundary conditions.}
    \label{tab:bendingBeamUniformTransverseLoadingBoundaryConditions}
\end{table}

\par
The polynomial ansatz for the Airy stress function used in the reference solution is:
\begin{equation}
    \airy^{\text{ref}}\lp x,y\rp = A_{20} x^2 +  A_{21} x^2y +  A_{03} y^3 +  A_{23} x^2y^3 - \frac{ A_{23}}{5}y^5,
    \label{eq:airyRef}
\end{equation}
where $A_{20}$, $A_{21}$, $A_{03}$, and $A_{23}$ are constant coefficients. These coefficients can be determined using the boundary conditions and the biharmonic equation such that the stress components are given as:
\begin{align}
    \stressTensorComponentRef{xx} &= \frac{3w}{4c}\lp\frac{l^2}{c^2}-\frac{2}{5}\rp y - \frac{3w}{4c^3}\lp x^2y-\frac{2}{3}y^3\rp,\\
    \stressTensorComponentRef{yy} &= -\frac{w}{2} + \frac{3w}{4c}y-\frac{w}{4c^3}y^3,\\
    \stressTensorComponentRef{xy} &= -\frac{3w}{4c}x + \frac{3w}{4c^3}xy^2.
\end{align}
\bigskip\par
\begin{figure}
    \centering
    \includegraphics[width=\textwidth]{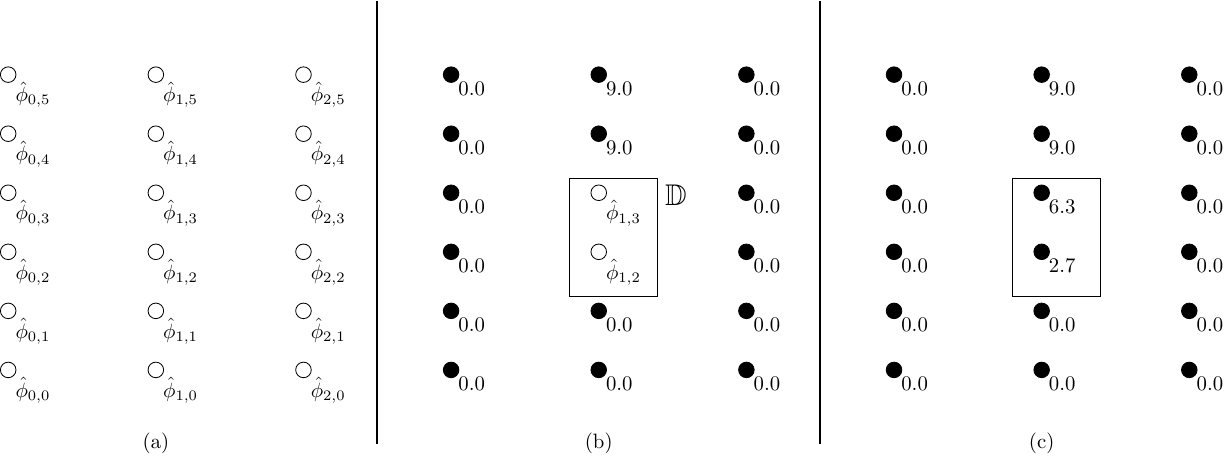}
    \caption{Bending of a beam by uniform transverse loading: control net for $\airyRef$ (a) initially, (b) after integration of the boundary conditions, and (c) after minimization of the complementary energy.
    \label{fig:bendingBeamUniformTransverseLoadingControlNet}}
\end{figure}

To derive the spline-based solution, we begin with defining the geometric mapping $\mapping$ as follows:
\begin{equation}
    \lp
    \begin{array}{c}
        x\\
        y
    \end{array}
    \rp
    =
    \mapping\lp\xi,\eta\rp
    =
    \lp
    \begin{array}{c}
        \lp2\xi - 1\rp l\\
        \lp1-2\eta \rp c
    \end{array}
    \rp.
\end{equation}
Next, we specify the ansatz for the stress function $\airyRef$.
For the B-spline basis functions, we choose the polynomial degrees $p=2$ and $q=5$ to match the highest exponents in $x$ and $y$ appearing in the reference Airy stress function $\airy^{\text{ref}}$ from \Cref{eq:airyRef}. 
The \deleted{grid of control variables $\cv_{i,j}$, the so-called} control net\deleted{,} is defined using $n=3$ and $m=6$ control variables, which represents the minimum number required to support the chosen degrees. It is illustrated in \Cref{fig:bendingBeamUniformTransverseLoadingControlNet}a. 
Note that the total number of unknowns \deleted{in our ansatz, i.e., the number of control variables,} is only $\nUnknowns =n\cdot m = 18$.
\deleted{In general, the polynomial degrees and the number of control variables can be selected based on the expected structure of the solution in a given problem or determined empirically to achieve a desired level of accuracy, while keeping them as low as possible to maintain computational efficiency.}
\par
\begin{table}
    \centering
    \begin{tabular}{lccc}
        \toprule
                & Horizontal  & Vertical & Moment\\
        \midrule
        \rule{0pt}{20pt}
        Top     & $\int_{-l}^{l} \stressTensorComponent{xy}^2\,dx$  & $\int_{-l}^{l} \lp\stressTensorComponent{y}-w\rp^2\,dx $ &-- \\
        \rule{0pt}{20pt}
        Bottom  & $\int_{-l}^{l} \stressTensorComponent{xy}^2\,dx$  & $\int_{-l}^{l} \stressTensorComponent{y}^2\,dx $          &--\\
        \rule{0pt}{20pt}
        Left    & $\int_{-c}^{c} \stressTensorComponent{x}^2\,dy$   & $\lp\int_{-c}^{c} \stressTensorComponent{xy}\,dy-wl \rp^2$& $\lp\int_{-c}^{c} \stressTensorComponent{x}y\,dy\rp^2$\\
        \rule{0pt}{20pt}
        Right   & $\int_{-c}^{c} \stressTensorComponent{x}^2\,dy$   & $\lp\int_{-c}^{c} \stressTensorComponent{xy}\,dy+wl \rp^2$&$\lp\int_{-c}^{c} \stressTensorComponent{x}y\,dy\rp^2$\\
        \bottomrule
    \end{tabular}
    \caption{Bending of a beam by uniform transverse loading: quantities to be minimized for a weak enforcement of the boundary conditions.}
    \label{tab:bendingBeamUniformTransverseLoadingWeakBoundaryConditions}
\end{table}
For the weak enforcement of the boundary conditions, we minimize the quantities provided in \Cref{tab:bendingBeamUniformTransverseLoadingWeakBoundaryConditions}.
After integration of the boundary conditions, only two control variables remain unconstrained, i.e., $|\setFreeControlVars|=2$ (see \Cref{fig:bendingBeamUniformTransverseLoadingControlNet}b), and are the \glspl{dof} for the minimization of the complementary energy.
In this case, we use the optimality condition and solve a linear system of equations for the remaining control variables $\cv_{i,j}\in\setFreeControlVars$. The resulting values for the control variables are shown in \Cref{fig:bendingBeamUniformTransverseLoadingControlNet}c.
\par
\begin{figure}
    \centering
    \includegraphics[width=\textwidth]{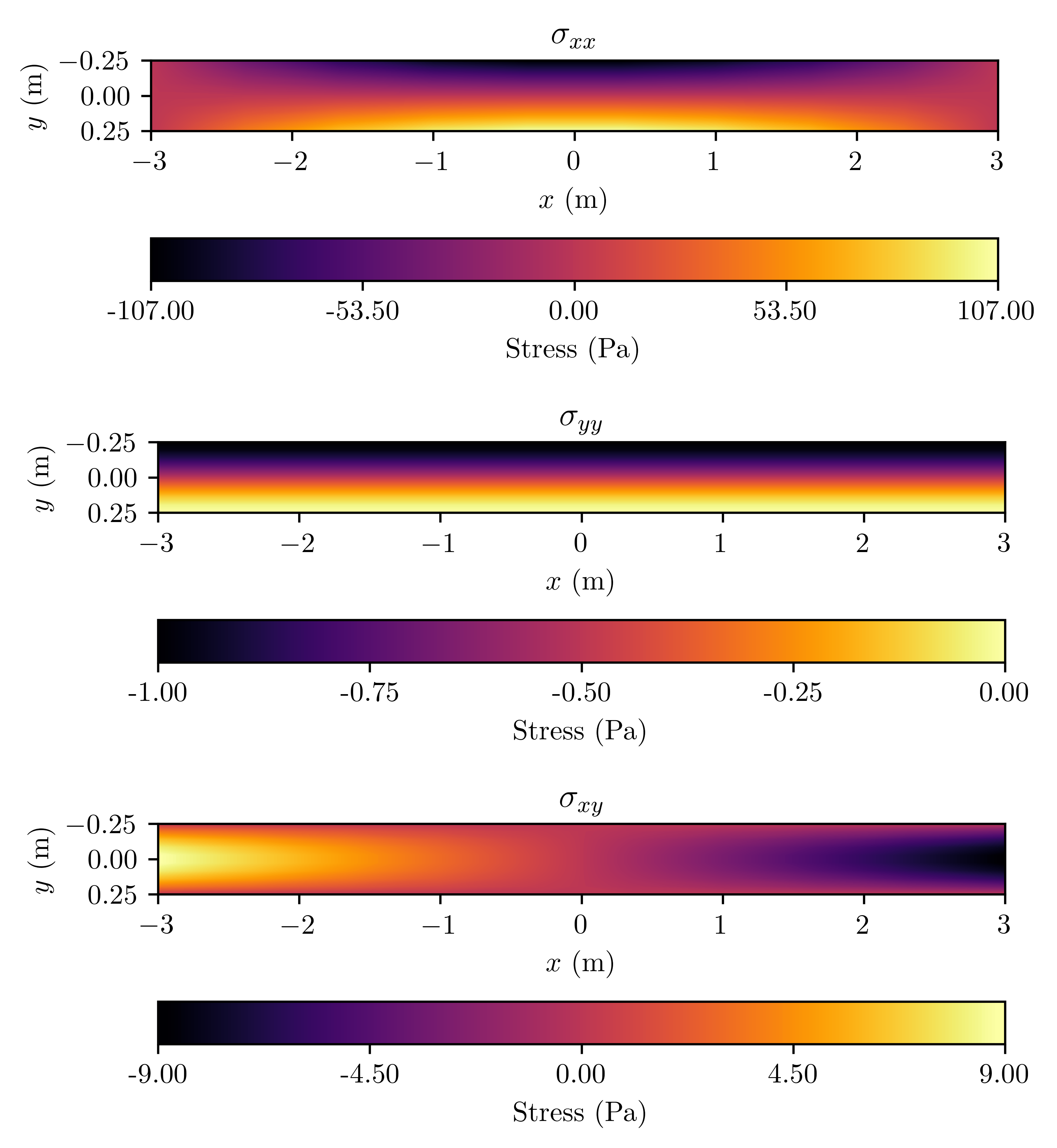}
    \caption{Bending of a beam by uniform transverse loading: stress distributions for the three stress components.}
    \label{fig:bendingBeamUniformTransverseLoadingStress}
\end{figure}

Finally, we can evaluate the second-order derivatives of $\airy$ to obtain the stress components in the physical space. 
The resulting distributions of the individual stress components are shown in
\Cref{fig:bendingBeamUniformTransverseLoadingStress}.
As can be seen in these plots, the spline-based approach yields a smooth and continuous stress distribution throughout the entire domain, despite the relatively small number of unknowns. This demonstrates the effectiveness of the high-order continuity inherent in the spline representation, in contrast to the inter-element discontinuities that typically arise in standard \glspl{fem}.
\bigskip\par
\begin{figure}
    \centering
    \includegraphics{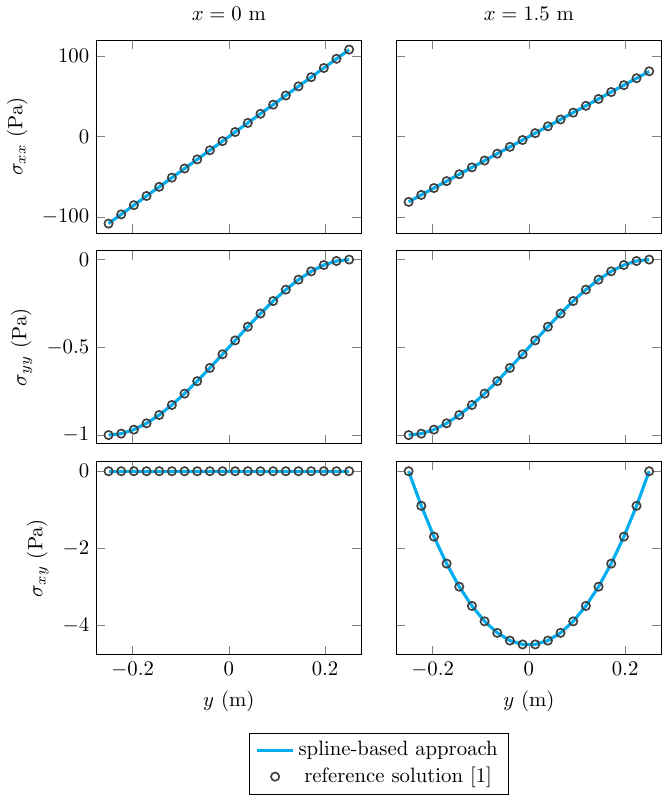}
    \caption{Bending of a beam by uniform transverse loading: comparison of the stress components across the beam’s height at the mid-span (left column, $x = 0~\si{\meter}$) and further along the beam (right column, $x = 1.5~\si{\meter}$).}
    \label{fig:bendingBeamUniformTransverseLoadingComparison}
\end{figure}

To validate our results, we compare them with the reference solution obtained from the power series approach presented in~\cite[Example~8.3]{Sadd2021}. Specifically, we examine the distribution of the individual stress components across the beam’s height at two locations: the mid-span ($x = 0~\si{\meter}$) and closer to the right end ($x =1.5~\si{\meter}$), as shown in \Cref{fig:bendingBeamUniformTransverseLoadingComparison}. Across all components, the spline-based solution exhibits excellent agreement with the reference, confirming the accuracy of the proposed method.
\par
For a quantitative comparison, we consider the relative difference between the spline-based solution and the reference solution in the $L_2$ norm over the entire domain, given as
\begin{equation}
    \relError{\stressTensorComponent{\bullet}} 
    = 
    \lp\frac{\intDomain{|\stressTensorComponent{\bullet}-\stressTensorComponent{\bullet}^{\text{ref}}|^2}}{\intDomain{|\stressTensorComponent{\bullet}^{\text{ref}}|^2}}\rp^{\frac{1}{2}}, 
\end{equation}
where $\stressTensorComponent{\bullet}^{\text{ref}}$ denotes the stress components obtained from the power series approach $\airyRef^{\text{ref}}$ used as the reference.
The resulting relative differences for the components $\stressTensorComponent{xx}$, $\stressTensorComponent{yy}$, and $\stressTensorComponent{xy}$ are listed in \Cref{tab:bendingBeamUniformTransverseLoadingRelError}, and amount to $1.19 \times 10^{-3}$, $2.99 \times 10^{-4}$, and $1.80\times10^{-4}$, respectively.
These small values confirm that the spline-based approach provides highly accurate stress predictions, even with a compact representation involving only a small number of control variables.
\begin{table}
    \centering
    \begin{tabular}{cccc}
        \toprule
         $l/c$&$\stressTensorComponent{xx}$& $\stressTensorComponent{yy}$ & $\stressTensorComponent{xy}$  \\
         \midrule
         $12$&$1.19\times10^{-3}$&  $4.91\times10^{-5}$ & $1.80\times10^{-4}$ \\
         $24$&$2.99\times10^{-4}$&  $1.20\times10^{-5}$ & $4.43\times10^{-5}$ \\
         $48$&$7.47\times10^{-5}$&  $3.00\times10^{-6}$ & $1.11\times10^{-5}$ \\
         \bottomrule
    \end{tabular}
    \caption{Bending of a beam by uniform transverse loading: relative differences $\relError{\stressTensorComponent{\bullet}}$ between the spline-based and reference solutions for various aspect ratios $l/c$.}
    \label{tab:bendingBeamUniformTransverseLoadingRelError}
\end{table}
The observed differences between the spline-based solution and the polynomial reference solution can be primarily attributed to boundary effects. In our approach, boundary conditions are enforced entirely in a weak form, whereas the reference solution uses a limited, carefully chosen set of polynomial terms designed to satisfy the simplified boundary conditions more directly---``a choice that has come from previous trial and error''~\cite[Example~8.3]{Sadd2021}.
In contrast, the spline-based ansatz inherently includes a much broader range of polynomial terms, offering greater flexibility but potentially requiring additional constraints to better reproduce the boundary behavior seen in the reference solution.
It is important to note, however, that the reference itself uses a simplification of the true boundary conditions. 
Therefore, attempting to exactly match the reference solution’s behavior at the boundaries is not meaningful. More importantly, the expected decrease in boundary effects with increasing beam aspect ratio $l/c$, as predicted by the Saint-Venant principle, should be verified, and is indeed confirmed by the data presented in \Cref{tab:bendingBeamUniformTransverseLoadingRelError}.
\subsection{Bi-Layer Cantilever with Anisotropic Material Behavior}
\label{subsec:biLayerCantilever}
After validating our method, we now turn to more challenging scenarios where accurate stress prediction is particularly demanding. In this section, we consider a multilayer planar beam with anisotropic material behavior under the plane stress assumption, as introduced in~\cite{Balduzzi2019}. This test case is especially relevant for modeling timber and composite structures, where layered configurations and directional material properties play a critical role.
Specifically, we investigate the stress prediction for a bi-layer anisotropic cantilever and perform a comparison with \gls{fem} results reported in the literature (see~\cite[Section~5.1]{Balduzzi2019}).
\par
\begin{figure}
    \centering
    \includegraphics[width=\textwidth]{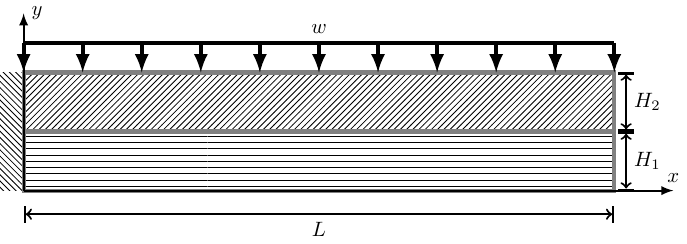}
    \caption{Bi-layer cantilever: setup.}
    \label{fig:bilayerCantileverSetup}
\end{figure}
The setup of the problem is illustrated in \Cref{fig:bilayerCantileverSetup}. The beam, which has a length $L$ and a unit depth, consists of two layers with thicknesses $H_1$ and $H_2$, respectively, and is subjected to a uniformly distributed loading $w$. 
Each layer exhibits piecewise constant material properties and the anisotropy is described through a position-dependent compliance tensor in matrix form:
\begin{equation}
    \complianceMatrix(y)
    =
    \rotationMatrix^{\text{T}}(y)
    \lb
    \begin{array}{ccc}
         \frac{1}{E_{11}(y)}& -\frac{\nu(y)}{E_{11}(y)} & 0 \\
         -\frac{\nu(y)}{E_{11}(y)}& \frac{1}{E_{22}(y)} & 0 \\
         0& 0 & \frac{1}{G_{12}(y)}
    \end{array}
    \rb
    \rotationMatrix(y),
\end{equation}
where $E_{11}(y)$, $E_{22}(y)$, $G_{12}(y)$, and $\nu(y)$ define the material properties in the local principal directions.
The rotation matrix
\begin{equation}
    \rotationMatrix(y)
    =
    \lb
    \begin{array}{ccc}
         \cos^2(\theta(y))& \sin^2(\theta(y)) &  \sin(2\theta(y))  \\
         \sin^2(\theta(y))& \cos^2(\theta(y)) & -\sin(2\theta(y))  \\
         -\frac{\sin(2\theta(y))}{2}& \frac{\sin(2\theta(y))}{2} & \cos(2\theta(y))
    \end{array}
    \rb,
\end{equation}
accounts for the rotation of the local principal axes by an angle $\theta(y)$ relative to the global $x$-axis.
In this example, the material principal direction is aligned with the beam axis in the bottom layer ($\theta = 0\degree$), while in the top layer it is rotated by $\theta = 15\degree$. The values for all relevant dimensions and material parameters are summarized in \Cref{tab:bilayerCantileverSettings}. 
\begin{table}
    \centering
    \begin{tabular}{cccccccc}
        \toprule
         $L$  & $H_1$ & $H_2$ & $w$ & $E_{11}$ & $E_{22}$& $G_{12}$ &$\nu$\\
         (\si{\milli\meter}) & (\si{\milli\meter}) & (\si{\milli\meter}) & (\si{\newton\per\milli\meter}) & (\si{\pascal}) & (\si{\pascal})&  (\si{\pascal})& (--)  \\
         \midrule
         $500$ & $50$ & $50$ & $1$ & $10\times10^9$ & $0.5\times10^9$ & $1\times10^9$ & $0$\\
         \bottomrule
    \end{tabular}
    \caption{Bi-layer cantilever: test case settings.}
    \label{tab:bilayerCantileverSettings}
\end{table}
\bigskip\par
Next, we proceed with the spline-based approach. We reflect the division of the beam into two layers by introducing separate patches and stress functions for each region, allowing us to capture the distinct material behavior in the multilayer configuration. Specifically, we define individual stress functions, $\airyRef^{\text{bottom}}$ and $\airyRef^{\text{top}}$, along with their respective geometric mappings:
\begin{align}
    \mapping^{\text{bottom}}(\xi,\eta) = 
    \lp
    \begin{array}{c}
         \xi L  \\
         \eta H_1
    \end{array}
    \rp,
    \quad
    \mapping^{\text{top}}(\xi,\eta) = 
    \lp
    \begin{array}{c}
         \xi L  \\
         H_1 + \eta H_2 
    \end{array}
    \rp.
\end{align}
We employ the same B-spline ansatz for both layers, using degrees $p=2$ and $q=4$, and control net dimensions $n=12$ and $m=7$, which results in a total of $\nUnknowns = 2\cdot n \cdot m = 168$ control variables.
Traction boundary conditions are imposed along the top edge to realize the external loading, and zero-traction conditions are enforced on the right edge and the bottom surface. No traction is specified on the left edge, as the zero-displacement boundary condition for the clamp automatically enters the formulation through the total complementary energy, which will be minimized.
To couple the two patches along the horizontal interface at $y=H_1$, we impose a coupling condition requiring the traction on either side of the internal boundary to be equal in magnitude but opposite in direction. All applied weak boundary and coupling conditions are summarized in \Cref{tab:bilayerCantileverWeakBoundaryConditions}.
\begin{table}
    \centering
    \begin{tabular}{llcc}
        \toprule
                & & Horizontal  & Vertical \\
        \midrule
        \rule{0pt}{20pt}
        \multirow{2}{*}{Top Layer} & Top     & $\int_{0}^{L} \left|\traction_x\right|^2\,dx$  & $\int_{0}^{L} \left|\traction_y+w\right|^2\,dx $  \\
        \rule{0pt}{20pt}
        &Right   & \multicolumn{2}{c}{$\int_{H_1}^{H_2} \left|\traction_i\right|^2\,dy$} \\
        \midrule
        \multicolumn{2}{l}{Coupling} & \multicolumn{2}{c}{$\int_{0}^{L} \left|\traction^{\text{bottom}}_i+\traction^{\text{top}}_i\right|^2\,dx$} \\
        \midrule
        \rule{0pt}{20pt}        
        \multirow{2}{*}{Bottom Layer} &Bottom  & \multicolumn{2}{c}{$\int_{0}^{L} \left|\traction_i\right|^2\,dx$} \\
        \rule{0pt}{20pt}
        &Right   & \multicolumn{2}{c}{$\int_{0}^{H_1} \left|\traction_i\right|^2\,dy$}\\
        \bottomrule
    \end{tabular}
    \caption{Bi-layer cantilever: quantities to be minimized for a weak enforcement of the boundary and coupling conditions.}
    \label{tab:bilayerCantileverWeakBoundaryConditions}
\end{table}
\bigskip\par
\begin{figure}
    \centering
    \includegraphics[width=0.9\textwidth]{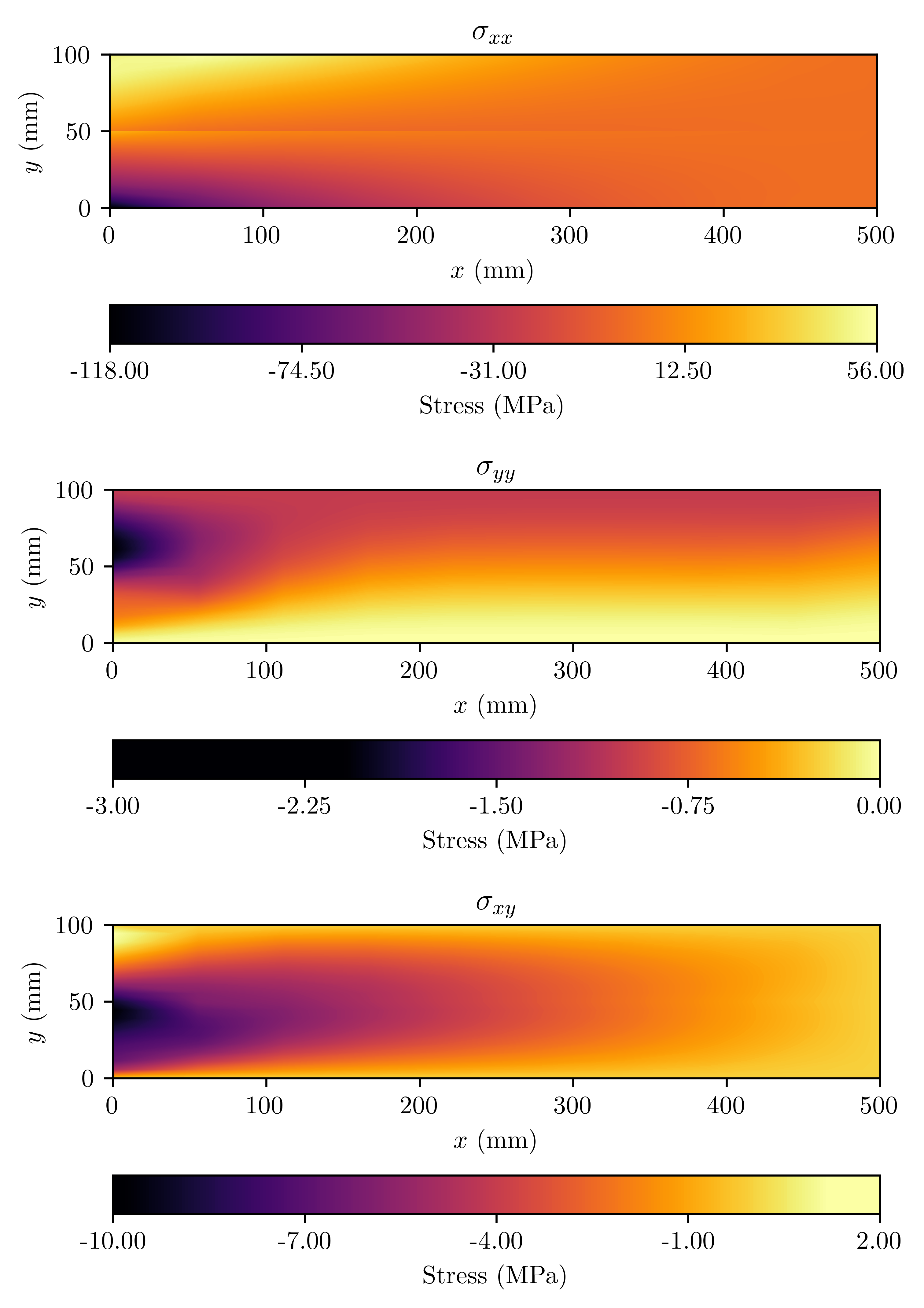}
    \caption{Bi-layer cantilever: stress distributions for the three stress components.}
    \label{fig:bilayerCantileverStress}
\end{figure}

The resulting stress distributions for all three components are shown in 
\Cref{fig:bilayerCantileverStress}.
Due to the anisotropic material behavior, the problem gives rise to non-trivial stress distributions, even within the geometrically simple rectangular domain. As seen in the figures, the stress fields are still smooth within each individual layer. Furthermore, the components relevant for the interface coupling, $\stressTensorComponent{yy}$ and $\stressTensorComponent{xy}$, are continuous across the interface, confirming the correct enforcement of the coupling. 
In contrast, the normal stress component $\stressTensorComponent{xx}$ exhibits a clear jump at the interface, which is consistent with the discontinuous material behavior between the two layers.
\bigskip\par
To further assess the accuracy of our method, we compare the results to a reference \gls{fem} solution presented in~\cite[Section~5.1]{Balduzzi2019}, which was computed using Abaqus with 4-node plane stress, fully integrated, bilinear elements (element code CPS4). The elements were arranged in a structured mesh with a characteristic element size of $\delta^{\text{ref}} = 0.25\,\mathrm{mm}$, determined through a mesh convergence study. The resulting number of unknowns can thus be estimated as $\nUnknowns^{\text{ref}} \approx 1,600,000$.
In contrast, our spline-based model employs $\nUnknowns= 168$ unknowns, which corresponds to a difference of roughly four order of magnitude in the number of unknowns (see \Cref{tab:bilayerCantileverUnknowns}).
\deleted{and a characteristic element length of $\delta = \sqrt{\Delta A} \approx 29\,\mathrm{mm}$.
Compared to the element length used in the reference with $\delta^{\text{ref}} = 0.25\,\mathrm{mm}$, our characteristic element length is roughly two orders of magnitude larger. Since the number of elements in a structured mesh grows quadratically with the inverse of the element size, this corresponds to a difference of roughly four orders of magnitude in the number of unknowns.} 
We will show that despite this drastic reduction, our spline-based approach still yields stress predictions in excellent agreement with the \gls{fem} solution.
\begin{table}[]
    \centering
    \begin{tabular}{ccc}
        \toprule
         2D \gls{fem}~\cite{Balduzzi2019} & Spline-Based Approach  & Reduction Factor\\
         \midrule
         $\nUnknowns_{\text{ref}}\approx 1,600,000$ & $\nUnknowns = 168$ & $\sim 10^4$\\
         \bottomrule
    \end{tabular}
    \caption{Bi-layer cantilever: number of unknowns for the reference \gls{fem} results and the spline-based approach, together with the associated reduction factor.}
    \label{tab:bilayerCantileverUnknowns}
\end{table}
\par
\begin{figure}
    \centering
    \includegraphics{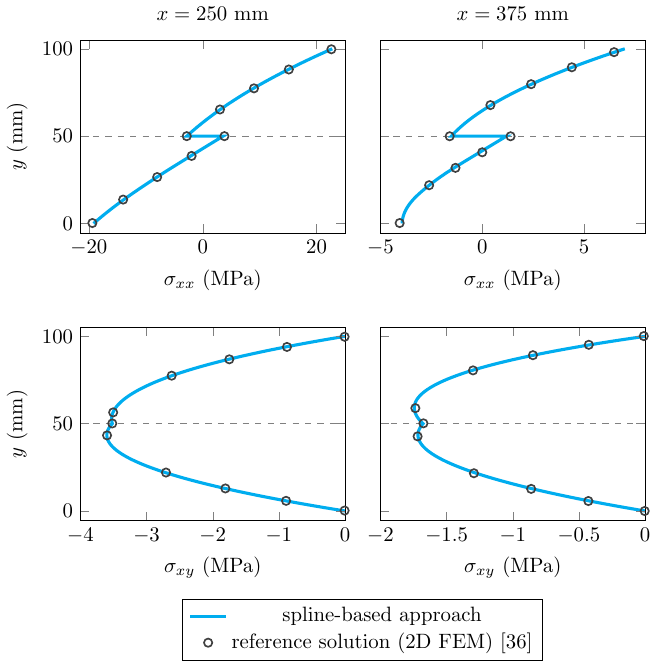}
    \caption{Bi-layer cantilever: comparison of the axial stress $\stressTensorComponent{xx}$ (top row) and the shear stress $\stressTensorComponent{xy}$ (bottom row) evaluated across the beam height at two positions: $x = 250\,\mathrm{mm}$ (left column) and $x = 375\,\mathrm{mm}$ (right column).}
    \label{fig:bilayerCantileverComparison}
\end{figure}

The comparison focuses on the axial stress $\stressTensorComponent{xx}$ and the shear stress $\stressTensorComponent{xy}$ evaluated across the beam height at two positions along the beam: $x = 250\,\mathrm{mm}$ and $x = 375\,\mathrm{mm}$. As shown in \Cref{fig:bilayerCantileverComparison}, the axial stress exhibits a clear discontinuity at the material interface, which is also accurately captured by the spline-based approach. The shear stress, in contrast, shows an asymmetric yet continuous profile that again matches well with the \gls{fem} results.
These results demonstrate that the spline-based formulation can deliver accurate stress predictions with a drastically reduced number of \glspl{dof}, highlighting its efficiency even in complex scenarios. 
\subsection{Parabolic-Shaped Cantilever}
\label{subsec:parabolicShapedCantilever}
As a final test case, we consider a non-symmetric cantilever with a parabolic profile, leading to a non-prismatic cross section. This example serves to demonstrate the applicability of our approach to problems beyond geometrically regular, rectangular domains. The specific test case, assuming plane stress conditions, was originally introduced in~\cite[Section~4]{Beltempo2015} and further examined in~\cite[Section~3.2]{Mercuri2020} as a representative example of non-prismatic structural elements, which are commonly used in civil engineering applications.
\par
\begin{figure}
    \centering
    \includegraphics[width=\textwidth]{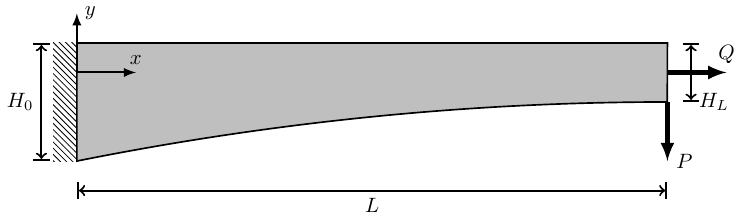}
    \caption{Parabolic-shaped cantilever: setup.}
    \label{fig:parabolicShapedCantileverSetup}
\end{figure}
The dimensions of the beam are illustrated in \Cref{fig:parabolicShapedCantileverSetup}. Over the beam length $L$, the height varies from $H_0$ at the fixed end to $H_L=H_0/2$ at the free end. While the upper edge remains horizontal, the lower edge follows a parabolic curve, resulting in an asymmetric, non-prismatic geometry. 
As before, it is assumed that the beam has a unit depth. 
At the right end, a horizontal force $Q$ and a vertical force $P$ are applied. The specific values for the geometric dimensions, loading conditions, and material parameters are listed in \Cref{tab:parabolicShapedCantileverSettings}.
\begin{table}
    \centering
    \begin{tabular}{ccccccc}
        \toprule
         $L$ (\si{\meter}) & $H_0$ (\si{\meter}) & $H_L$ (\si{\meter})& $Q$ (\si{\kilo\newton}) & $P$ (\si{\kilo\newton}) & $E$ (\si{\pascal}) &$\nu$ (--)  \\
         \midrule
         $5$ & $1$ & $0.5$ & $100$ & $-100$ & $1\times10^5$ & $0.3$\\
         \bottomrule
    \end{tabular}
    \caption{Parabolic-shaped cantilever: test case settings.}
    \label{tab:parabolicShapedCantileverSettings}
\end{table}
\bigskip\par
For the spline-based approach, the resulting geometric mapping is given by
\begin{equation}
    \mapping(\xi,\eta) 
    =
    \lp
    \begin{array}{c}
         \xi L  \\
         \frac{H_0}{4}\lp2\eta +\lp1-\eta\rp\lp-2\xi^2+4\xi-2\rp-1\rp 
    \end{array}
    \rp.
\end{equation}
We choose polynomial degrees $p=6$ and $q=4$, and define a control net consisting of $n=10$ by $m=5$ control variables, yielding a total of $\nUnknowns=50$.
Along the top and bottom edges, zero-traction conditions are applied, while on the right edge the loading is enforced through traction conditions with equivalent resultant forces. As before, the zero-displacement condition on the left edge is implicitly enforced through the total complementary energy and requires no additional treatment. An overview of the weakly enforced boundary conditions is provided in \Cref{tab:parabolicShapedCantileverWeakBoundaryConditions}.
\begin{table}
    \centering
    \begin{tabular}{lcc}
        \toprule
                & Horizontal  & Vertical \\
        \midrule
        \rule{0pt}{20pt}
        Top     & \multicolumn{2}{c}{$\int_{0}^{L} \left|\traction_i\right|^2\,dx$} \\
        \rule{0pt}{20pt}
        Bottom  & \multicolumn{2}{c}{$\int_{0}^{L} \left|\traction_i\right|^2\,dx$} \\
        \rule{0pt}{20pt}
        Right   & $\lp\int_{0}^{H_L} \traction_{x}\,dy -Q\rp^2$   & $\lp\int_{0}^{H_L} \traction_{y}\,dy -P\rp^2$ \\
        \bottomrule
    \end{tabular}
    \caption{Parabolic-shaped cantilever: quantities to be minimized for a weak enforcement of the boundary conditions.}
    \label{tab:parabolicShapedCantileverWeakBoundaryConditions}
\end{table}
\bigskip\par
\begin{figure}
    \centering
    \includegraphics[width=0.95\textwidth]{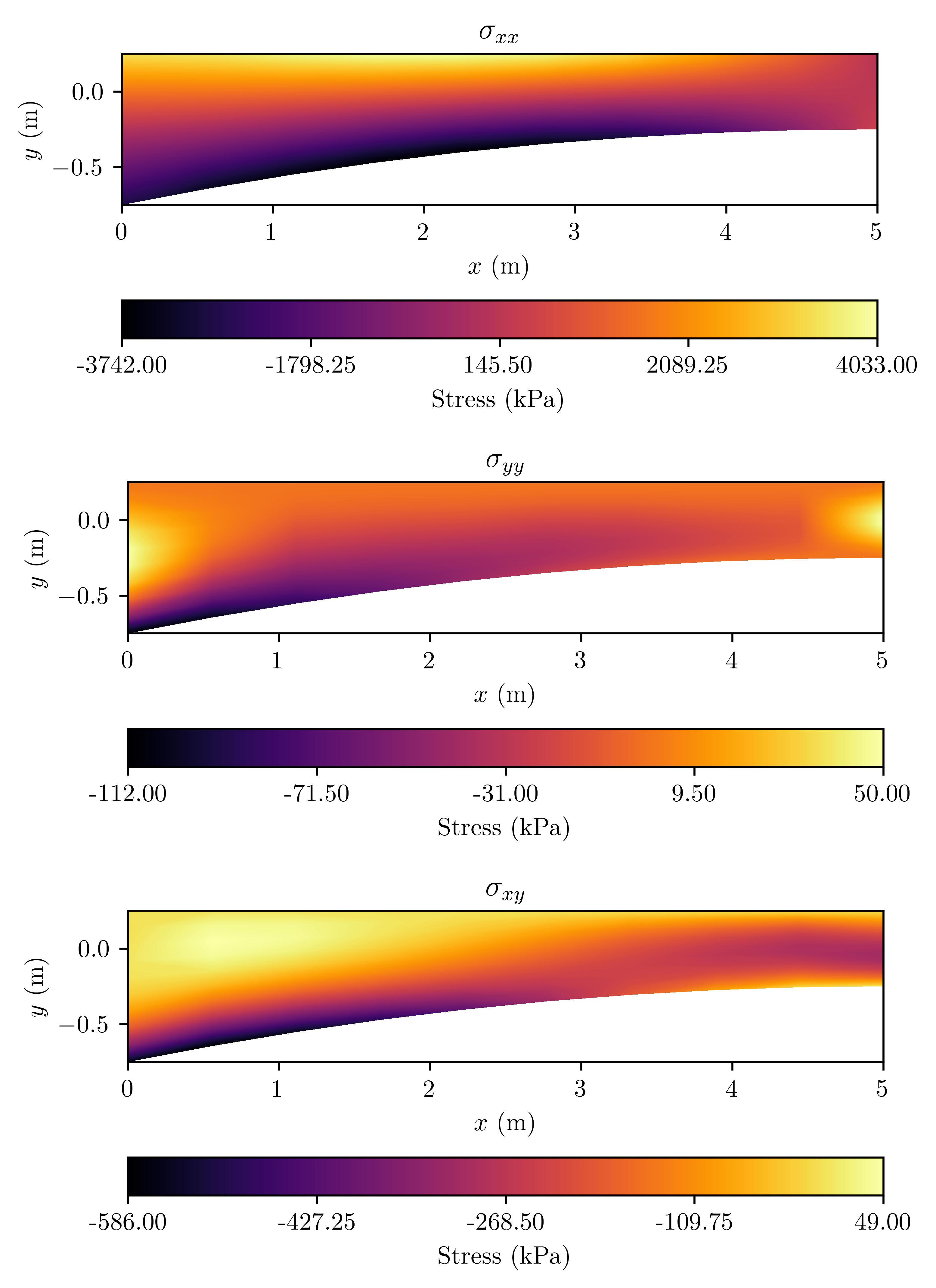}
    \caption{Parabolic-shaped cantilever: stress distributions for the three stress components.}
    \label{fig:parabolicShapedCantileverStress}
\end{figure}

\Cref{fig:parabolicShapedCantileverStress} presents
the resulting stress distributions for the three stress components. Due to the non-prismatic geometry and the combined loading conditions, the stress fields exhibit complex, non-trivial patterns. Despite this complexity, the spline-based formulation yields smooth and coherent stress fields. As in the previous example, we assess their accuracy by comparison with a reference \gls{fem} solution reported in the literature~\cite[Section~3.2]{Mercuri2020}. 
The reference solution was computed using Abaqus with a structured mesh of $2000 \times 150$ 4-node bilinear plane stress elements (element type CPS4).
Consequently, we estimate the resulting number of unknowns as $\nUnknowns^{\text{ref}} \approx 600,000$. In contrast, the spline-based approach uses only $\nUnknowns = 50$ unknowns, representing a reduction of four orders of magnitude (see \Cref{tab:parabolicShapedCantileverUnknowns}). This aligns with the efficiency gains observed in the previous test case. 
\begin{table}[]
    \centering
    \begin{tabular}{ccc}
        \toprule
         2D \gls{fem}~\cite{Mercuri2020} & Spline-Based Approach  & Reduction Factor\\
         \midrule
         $\nUnknowns_{\text{ref}}\approx 600,000$ & $\nUnknowns = 50$ & $\sim 10^4$\\
         \bottomrule
    \end{tabular}
    \caption{Parabolic-shaped cantilever: number of unknowns for the reference \gls{fem} results and the spline-based approach, together with the associated reduction factor.}
    \label{tab:parabolicShapedCantileverUnknowns}
\end{table}
As before, we now assess whether the spline-based formulation can still provide accurate stress predictions when compared to the detailed \gls{fem} reference solution.
\par
\begin{figure}
    \centering
    \includegraphics{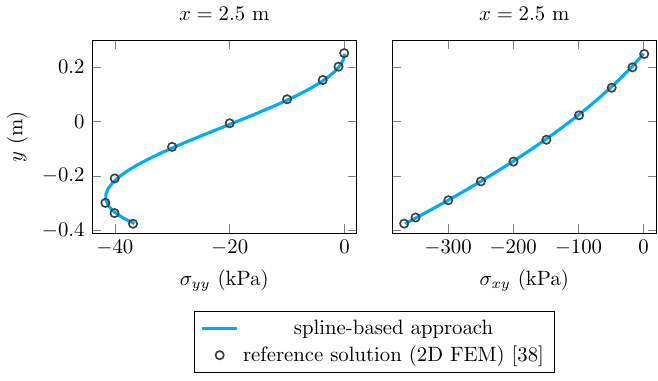}
    \caption{Parabolic-shaped cantilever: comparison of the transverse normal stress $\stressTensorComponent{yy}$ (left) and the shear stress $\stressTensorComponent{xy}$ (right) across the beam’s height at the mid-span ($x = 2.5~\si{\meter}$).}
    \label{fig:parabolicShapedCantileverComparison}
\end{figure}

Specifically, we compare the transverse normal stress $\stressTensorComponent{yy}$ and the shear stress $\stressTensorComponent{xy}$ at the mid-span of the beam ($x = 2.5~\si{\meter}$). Their distribution across the beam’s height is shown in \Cref{fig:parabolicShapedCantileverComparison}. The comparison demonstrates excellent agreement between the spline-based solution and the \gls{fem} reference, with the spline-based approach accurately capturing the overall stress profiles but also localized features like the inflection point in $\stressTensorComponent{yy}$.
This highlights the effectiveness of the proposed approach also in geometrically more complex domains.
\section{Conclusion and Outlook}
\label{sec:conclusion}
In this work, we presented a spline-based formulation for stress prediction in two-dimensional linear elasticity problems. The method is based on a B-spline discretization of the Airy stress function over a parametric domain. By introducing one or multiple geometric mappings, complex physical domains can be effectively represented. Thanks to the characteristics of the B-spline basis functions and their derivatives, the incorporation of traction boundary conditions and additional constraints is greatly simplified. The resulting stress fields are obtained through a variational approach, specifically the principle of minimum total complementary energy.
\bigskip\par
To assess the performance of the proposed approach, we conducted a series of numerical experiments, which demonstrated several key strengths of the method. First, validation against analytical solutions confirmed the accuracy of the spline-based formulation for classical benchmark problems including a bar under self-weight and the bending of a beam by uniform transverse loading.
\par
We then applied the method to more challenging scenarios involving either complex material behavior or non-trivial geometry. Specifically, we investigated (1) a bi-layer cantilever composed of anisotropic materials, and (2) a cantilever with a non-prismatic, parabolic profile. In both cases, comparisons with high-resolution \gls{fem} reference solutions from the literature showed excellent agreement in the stress fields. Notably, for these test cases, the spline-based approach achieved the same level of accuracy with a number of unknowns that was four orders of magnitude smaller than that of the reference models.
\par
Together, these results demonstrate that the proposed formulation effectively addresses the limitations of conventional methods: analytical approaches are often restricted to simple problems, while \gls{fem} approaches, either displacement- or stress-based, typically require a large number of unknowns to obtain accurate stress predictions.
\bigskip\par
Nonetheless, the approach also comes with certain limitations that define its current scope.
These limitations are rooted both in the underlying methodology, i.e., the use of the Airy stress function and the principle of minimum complementary energy, as well as in the chosen approximation framework. 
First, the formulation is restricted to linear elasticity to ensure that we can apply the principle of minimum complementary energy. While extensions to geometrically nonlinear settings have been investigated, such generalizations remain an active area of research, as discussed in~\cite{Santos2011}. 
Second, the assumption of plane problems arises directly from the use of the Airy stress function.
Finally, B-splines were used for the approximation due to the ease of implementation.
\bigskip\par
Despite these limitations, the presented formulation opens up promising directions for further research and application. While the present work focuses on plane problems, the framework is not restricted to two dimensions. A natural extension to three-dimensional elasticity could be achieved by adopting non-scalar stress functions, such as the Morera stress function, and applying the spline-based approximation component-wise. The remainder of the formulation, including the treatment of boundary conditions and stress evaluation, would remain essentially unchanged. Additionally, future work may also focus on enhancing the expressiveness of the stress representation by employing more general spline-based bases such as \gls{nurbs}.
These developments could significantly broaden the scope and applicability of the proposed approach.
\par
Beyond these potential extensions, the methodology is already relevant for ongoing developments in applied mechanics. For instance, a recent formulation of structural design optimization for \gls{qa}~\cite{Key2024} builds upon the principle of minimum complementary energy. Since \gls{qa} requires unconstrained minimization problems, the presented stress function-based approach, which automatically satisfies the constraint of static admissibility, offers a substantial benefit. Similarly, in the context of data-driven material identification using \glspl{pinn}~\cite{vanderHeijden2025}, the use of a stress function ensures internal equilibrium by construction, making it an attractive option for embedding physical consistency into the learning process.

\section*{CRediT authorship contribution statement}
\textbf{Fabian Key:} Conceptualization, Data Curation, Formal analysis, Investigation, Methodology, Software, Validation, Visualization, Writing – original draft, Writing – review \& editing
\textbf{Lukas Freinberger:} Software, Writing – review \& editing

\section*{Declaration of competing interest}
The authors declare that they have no known competing financial interests or personal relationships that could have appeared to influence the work reported in this paper.

\section*{Funding sources}
This research did not receive any specific grant from funding agencies in the public, commercial, or not-for-profit sectors.

\section*{Declaration of generative AI and AI-assisted technologies in the writing process}
During the preparation of this work the authors used ChatGPT in order to improve the clarity and fluency of the language. After using this tool/service, the authors reviewed and edited the content as needed and take full responsibility for the content of the published article.

\section*{Data availability}
The data generated in the course of this study is available from the TU Wien Research Data repository at \url{https://doi.org/10.48436/7sr7p-84y06}.





\bibliographystyle{elsarticle-num} 
\bibliography{references}






\end{document}